%% file: root.tex
\documentclass[review]{elsarticle}

\usepackage{lineno,hyperref}

\input{preamble}

\journal{Information Sciences}

\begin{document}

\begin{frontmatter}

\title{Privacy in Distributed Computations based on Real Number Secret Sharing} 

\author{Katrine Tjell\corref{mycorrespondingauthor}}
\ead{kst@es.aau.dk}
\author{Rafael Wisniewski}
\ead{raf@es.aau.dk}
\address{Department of Electronic Systems, Aalborg University, Fredrik Bajers Vej 7C, 9220 Aalborg, Denmark.}


\cortext[mycorrespondingauthor]{Corresponding author}


\begin{abstract}
Privacy preservation in distributed computations is an important subject as digitization and new technologies enable collection and storage of vast amounts of data, including private data belonging to individuals. To this end, there is a need for a privacy preserving computation framework that minimises the leak of private information during computations while being efficient enough for practical usage. This paper presents a step towards such a framework with the proposal of a \textit{real number secret sharing scheme} that works directly on real numbers without the need for conversion to integers which is the case in related schemes. The scheme offers computations like addition, multiplication, and division to be performed directly on secret shared data (the cipher text version of the data). Simulations show that the scheme is much more efficient in terms of accuracy than its counterpart version based on integers and finite field arithmetic. The drawback with the proposed scheme is that it is not perfectly secure. However, we provide a privacy analysis of the scheme, where we show that the leaked information can be upper bounded and asymptotically goes to zero. To demonstrate the scheme, we use it to perform Kalman filtering directly on secret shared data.
\end{abstract}


\end{frontmatter}


\section{Introduction}
In recent years, there has been a rapid development of technologies for digitization and collection and storage of data. Consequently, various distributed algorithms for the efficient processing of the collected data are being developed in many research communities like signal processing, control, machine learning, and optimization. Simultaneously, concerns about privacy and the possible misuse of the data means a sudden big interest in embedding cryptographic methods into the distributed algorithms to achieve privacy preserving data processing, \cite{SecControlHomo, janeAdditive, LCSS}.

So far, efficient data processing and privacy preservation are two terms that seems difficult to combine since the cryptographic methods tend to bring a substantial overhead in either communication, computation or both. Moreover, security of cryptographic methods such as secret sharing and homomorphic encryption relies on modular arithmetic, which entails that all data to be protected must be integers and computations on this data must be translated into equivalent computations using finite field arithmetic, \cite{PrivOptHomo, JaneShamir1, CDC19}.
The drawbacks of this are, for instance, loss of precision in the solution (because of rounding decimal numbers to integers) and that many operations such as division becomes very intractable.

For some applications, efficient processing, that is not constrained to finite field arithmetic, is crucial. Thus, it becomes relevant to consider a trade-off between privacy and efficiency since after all; limited privacy is better than none. To this end, we explore distributed computations in the secure multiparty computation \cite{cd} setup, where only cipher texts travel between participants and plain texts stay hidden throughout computations. Essentially, what we propose is a \textit{real number secret sharing scheme} that circumvents the disadvantages of using only integers and modular arithmetic and consequently achieves improved performance compared to state-of-the-art methods. The scheme works directly on real numbers and we show straight forward implementations of addition, multiplication and division performed directly on the secret shared data. The shortcoming to our proposed scheme is that it does not guarantee perfect security like its counterpart version based on integers and modular arithmetic. However, we carefully control the amount of leaked information and provide information theoretic results to support our claims. 

As a motivating example, we demonstrate the use of the proposed scheme to perform privacy preserving Kalman filtering. That is, we consider a linear dynamical system with state-transition matrix $\v A$, control input matrix $\v B$, control input $\v u_k$, process noise $\v w_k$ and state vector $\v x_k$: 
\begin{equation}
    \v x_k = \v A \v x_{k-1} + \v B \v u_{k} + \v w_k. \label{one}
\end{equation}
Observations (or measurements) of the state vector, $\v z_k$ are modeled as
\begin{equation}
    \v z_k = \v H \v x_{k} + \v v_k \label{two},
\end{equation}
where $\v H$ is the observation matrix and $\v v_k$ is the measurement noise. The objective is to estimate the true state of the system from the noisy observations, which is optimally done using the Kalman filter. The privacy concern emerges from the measurements which could be private data that potentially leaks private information. Scenarios where a problem of this form appears, could for instance be traffic monitoring \cite{KalTraffic}, medical monitoring \cite{KalHealth}, and consumption forecasting \cite{KalWaterForecast}. The problem of privacy preserving Kalman filtering has already been studied for instance in \cite{PrivAwareKal} that uses a form of data compression to preserve privacy of measurements, \cite{secKalman} that base the privacy on a combination of homomorphic encryption and secure multiparty computation techniques, and \cite{KalDiff} that relies on differential privacy. These existing works all suffer from a degradation in output utility compared to the none-privacy preserving solution due to noise insertion or to the previously mentioned rounding of reals to integers. We will show that a privacy aware Kalman filter based on our real number secret sharing scheme achieves significantly improved output utility. Furthermore, we compare our privacy preserving Kalman filter to the one proposed in \cite{secKalman} and show that ours has a reduction in computation and communication overhead. 

\subsection{State of the art}

The typical way of preserving privacy of real numbers is to simply discard the decimals and keep the integer part which is the suitable representation for most cryptographic methods, \cite{ecc,Darup2018,jane2019} . The induced error bounds caused by the truncation, can be made small by introducing scaling constants prior to truncation. However, the size of the modular field, in which the cryptographic calculations take place, increases according to the size of the scaling factors and thus cannot be made arbitrarily big.  

One of the first more direct ways to deal with non-integers in cryptographic computations, was made in 2010 by Catrina et al. in \cite{catrinaFixedPoint}. Their proposed solution builds on a fixed-point representation of real numbers that allows the use of Shamir's secret sharing scheme as the underlying cryptographic technique. In \cite{catrinaLinearProg} this solution was applied to privacy preserving linear programming. Along this line of research, \cite{Aliasgari2013} proposed in 2013 a similar secure floating-point computation scheme also based on a linear secret sharing framework. In 2016, \cite{Dimitrov2016} proposed other techniques for representing secure real numbers suitable for a secret sharing framework with their so-called golden-section and logarithmic number formats. 

Apart from secret sharing based secure computation frameworks, there has also been several attempts to secure real number computations in homomorphic encryption based frameworks, \cite{realNumHomo, realNumHomo2, realNumHomo3}. Analog to the approach based on secret sharing, the main idea here is to convert the real number into a multi-bit binary integer to achieve a fixed precision presentation of a real number. The drawback with these approaches is the time consuming computational overhead with homomorphic encryption and also that the proposed schemes only offer addition and in some cases multiplication of cipher-texts. This is in contrast to our scheme that allows addition, multiplication and division to be performed efficiently on the cipher-text data.

Finally, our work is closely related to \cite{infSecretSharing} that considers secret sharing schemes (SSS) over infinite domains, e.g. the real number line. Among others, they propose a scheme very similar to ours which is based on polynomials and Lagrange interpolation. However, they consider a game between a dealer and an adversary, which is for the dealer to chose a scheme and a secret such that the adversary has the least probability of guessing the secret. On the contrary, our work assumes that a group of parties would like to perform computations without exposing data belonging to the individual parties. In this sense, the secret is the data, and not something we can chose to our liking. Also, we provide a quantification of the privacy loss of the scheme and propose how to use the scheme for secure multiparty computation (SMPC), which \cite{infSecretSharing} does not.

\subsection{Contribution}
The paper puts forth a real number secret sharing scheme which bypasses the usual restrictions to integer secrets and finite field computations. This makes the scheme very practical as solutions can be calculated with high precision and without the need for computations being performed with modular arithmetic. The scheme performs the same or with significant less computation and communication complexity compared to state-of-the-art methods. In Table \ref{tab:comparison1} the number of interactive operations (IO) are given for a selected number of state-of-the-art protocols. IO's are those that require communication between the participants, and since the time spent on local computations vanishes compared to time spent on IO's, this measure gives both an indication of communication and computation complexity. 

\begin{table}
\centering
\caption{Comparison of interactive operations (IO) of state-of-the-art protocols, where $l_t$ is the bit-length of the truncated secret and $l$ and $k$ is, respectively, the bit-length of the significant and exponent of the fixed point represented secret.}
\label{tab:comparison1}
\resizebox{\textwidth}{!}{%
\begin{tabular}{lcccl}
\hline
                                                                                                                             & IO addition                                                                                                    & IO multiplication & IO division               & \multicolumn{1}{c}{Precision} \\ \hline
Shamir's SSS with truncation \cite{ecc}                                                                     & 0                                                                                                              & 2                 & $220l_t+ \log2 l_t + 238l_t+3$  & Up to scaling                 \\
\begin{tabular}[c]{@{}l@{}}Shamir's SSS with fixed point\\ representation \cite{Aliasgari2013}\end{tabular} & \begin{tabular}[c]{@{}c@{}}$14 l + 9k + (\log l) \log \log l$\\ $ + (l+9) \log l + 4 \log k + 37  $\end{tabular} & $8l+10$           & $2 \log l (l+2) + 3 l +8$ & Up to scaling                 \\
Real numbers SSS                                                                                                             & 0                                                                                                              & 2                 & 3                         & Machine precision             \\ \hline
\end{tabular}
}
\end{table}

The main contribution of the paper can be summarized as:
\begin{itemize}
\item To the best of our knowledge, this is the first attempt for a SMPC scheme that works directly on the real number line and consequently offers a trade-off between privacy and practicality.  
\item The proposed scheme bypasses the requirements for modular arithmetic and integer secrets which is in contrast to state-of-the-art SMPC techniques.
\item The scheme allows addition, multiplication and division to be performed directly on shares (ciphertext version of the data), opposed to related schemes that typically only allow addition and in some cases multiplication.
\item The paper provides an in-depth analysis of the privacy guaranties of the scheme as well as a quantification of leaked data.
\end{itemize}

\subsection{Outline}
The paper proceeds in section \ref{sec:prelim} by introducing the preliminaries and giving motivation for the work. Section \ref{sec:problem} states formally the problem of the paper, while section \ref{sec:proposed} presents the proposed scheme and the privacy analysis. In section \ref{sec:numerical_eval} we give a numerical evaluation of the proposed scheme, while section \ref{sec:application} provides simulations of the scheme for Kalman filtering and finally, section \ref{sec:conc} concludes the paper.

\section{Preliminaries and Motivation}\label{sec:prelim}
In this section, we clarify our notation and terminology and afterwards we give a brief introduction to the concept of \textit{secret sharing} and SMPC, while subsequently discussing their shortcomings which motivates the work in this paper.

\subsection{Notation and Terminology}

\begin{figure*}[t!]
    \centering
    \begin{subfigure}[t]{0.48\textwidth} 
        \centering
        \includegraphics[scale=1]{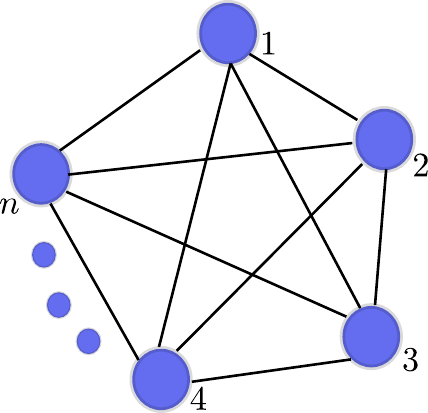} \caption{$n$ participants that can communicate privately with each other.}
        \label{fig:p1}
    \end{subfigure}%
    ~
    \begin{subfigure}[t]{0.48\textwidth} 
        \centering
        \includegraphics[scale=1]{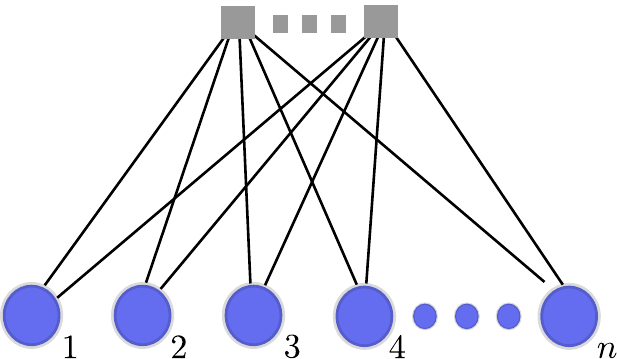}
        \caption{$n$ participants and \textit{computing parties} (the grey squares). }
        \label{fig:p2}
    \end{subfigure}
    
    \caption{Illustration of two scenarios of the communication network. The first scenario (a), each participant can communicate privately with each of the other participant, and all computations are performed by the participants themselves. In (b) each participant can communicate privately with a number of \textit{computing parties} (the grey squares) and each of the computing parties can communicate privately with each of the other computing parties. The computing parties receives shares of the input data from the participants and perform all computations without learning the secret data.}
    \label{fig:parties}
\end{figure*}

Let \p be an index set of $n > 2$ participants. We assume that each participant $p\in \p$ can communicate privately with each of the other participants $j \in \p$ or alternatively that there exists a number of computing parties that each participant can communicate with. Each of these scenarios is illustrated in \figref{fig:parties}. The advantage of the second scenario is that the computing parties do most of the computations and hence the participants do not have to possess large computation capabilities. Furthermore, the participants need only to communicate with a number of computing parties (which can be as low as 3). In the remaining of the paper we do not make a distinction between these two scenarios, but remark that any presented method can straightforward be used in both.

Concerning notation, let $s$ be a secret value belonging either to a participant or to an external entity providing secret data. We use $\SE{s}$ to denote the set of so-called \textit{shares} of $s$. In other words, each share $\se{s}$ is a cipher-text version of $s$. Combining a set $\TE{s}{p}$ of shares, for $\T \subseteq \p$ where $t < |\T| \leq n$ and $t$ is an integer threshold, the shares can be deciphered and $s$ recreated.

\subsection{Secret Sharing, SMPC and their shortcomings} \label{sec:shortcom}
Secret sharing in general lets a party "share" a secret among $n$ participants, such that at least $t+1 \leq n$ of the participants must cooperate to learn the secret and opposite; no subset of less than $t+1$ participants gets information about the secret. There are many different secret sharing schemes, each tailored to different use cases. Perhaps the most simple (and intuitive) secret sharing scheme is the \textit{additive} one \cite{cd}, where $t=n$ meaning that all shares are needed to reconstruct the secret. In this scheme, the shares $\SE{s}$ of the secret $s$ satisfy that
\begin{equation}\label{eq:additive}
    s = \br{\sum_{p\in\p} \se{s}} \mod q,
\end{equation}
where $q$ is a large prime number. When choosing $n-1$ of the shares uniformly on $[0,q-1]$ and the last share such that \eqref{eq:additive} holds, the modular arithmetic ensures that all shares are uniformly distributed. This means that the scheme is perfectly secure since the uniform distribution holds no information about the secret. The disadvantage is that $s$ must be an element of $\F$, where \F is a finite field of $q$ elements.  

Many secret sharing schemes, like the additive one and Shamirs scheme \cite{Shamir}, are very useful in SMPC protocols. These protocols, lets $n$ participants compute a function, that takes as input a private value from each participant, while keeping the private values secret. For instance, for secrets \m{s_1,s_2 \in \F,} the sum $s_1+s_2$ can be calculated directly on additive shares of each of the secrets;
\begin{equation}
    s_1 + s_2 = \br{ \sum_{p \in \p} \se{s_1}+\se{s_2}} \mod q,
\end{equation}
where \m{\se{s_1} + \se{s_2}} is computed by the $p$'th participant.

The drawback is that $q$ must be bigger than $s_1+s_2$ in order to get the correct result and if no information about the secret data is available, it can be difficult to choose $q$. 

More advanced schemes like Shamir's scheme, also allows multiplication of secrets directly on the shares and in principle also division. However, the division will be \textit{finite field division} \cite{justesen} and not real number division. As introduced in \cite{ecc}, there are complicated tricks, which usually involve bit-decomposition of the secrets, that will enable the computation of real number division performed on the shares. However, say that the secret to be divided is $-3$ (which would be represented as $q-3$ in \F), what effectively would happen is the division of $q-3$ and not $-3$, which would lead to incorrect results. This is an example of how finite field arithmetic complicates the computations which leads to part of our motivation to introduce a real number secret sharing scheme that does not depend on finite field arithmetic. 

\section{Problem Statement}\label{sec:problem}
Upon the discussion in section \ref{sec:shortcom}, we conclude that the problem of preserving privacy of real numbers without being limited to finite field arithmetic is indeed a relevant topic in privacy preserving computations. To address this problem, we will propose a real number secret sharing scheme. To this end, we start with the following definition. 

\begin{defi}[Real Number Secret Sharing Scheme]\label{defi:share}
	A real number secret sharing scheme consists of two algorithms; $\mathtt{share}$ and $\mathtt{recon}$. $\mathtt{share}(s,t,\p) = \SE{s}$ takes a secret $s\in \R$, the threshold $t\in \mathbb{N}$ with $t < n$ and the indices of $n$ participants $\p$ and outputs a share $\se{s} \in \R$ for each participant $p \in \p$. The algorithm $\mathtt{recon}(\TE{s}{p}) = s$ outputs the secret $s$ upon inputting at least $t+1$ shares from any set of participants $p \in \T$, where $\T \subseteq \p$ with $|\T| > t$.
\end{defi}
We have the following requirements for the real number secret sharing scheme. 
\begin{itemize}
    \item \textbf{Correctness}. A reconstructed secret should be equal to the original secret, that is $s - \mathtt{recon}(\TE{s}{p} = 0$.

    \item \textbf{Privacy.} Only by combining at least $t$ shares of $s$ should it be possible to reconstruct $s$. A set of fewer than $t$ shares should reveal only very little information about $s$. We state this formally by using the information theoretic measure called mutual information \cite[p.250]{Cover};
    \begin{equation}
        I(X; Y) = h(X) - h(X|Y), 
    \end{equation}
    where $h(X)$ is the entropy of the random variable $X$ and $h(X|Y)$ is the conditional entropy of $X$ given the random variable $Y$. The mutual information $I(X;Y)$ can be interpreted as the reduction in uncertainty about $X$ one has after learning the outcome of $Y$ (and vice versa since mutual information is symmetric). To this end, we use $S$ and $\se{S}$ to denote the random variables that has $s$ and $\se{s}$ as outcomes, and we require that for any $\delta > 0$ there exists $\TE[\T']{S}{p}$ such that
    \begin{equation}
        I(S; \TE[\T']{S}{p}) \leq \delta,
    \end{equation}
    where $\T' \subset \p$ with $|\T'| \leq t$. 
    \item \textbf{Computations directly on shares}. At least the operations addition, multiplication, and division, should be applicable directly on shares. That is, for any secrets $s_1,s_2 \in \R$ and properly defined protocols $\mathtt{add}, \mathtt{mult}$, and $\mathtt{inv}$, the following should hold 
    \begin{align}
        \mathtt{recon}( \tub{ \mathtt{add}( \se{s_1} , \se{s_2} )}_{p\in \T} ) &= s_1 + s_2 \\
        \mathtt{recon}( \tub{\mathtt{mult} (\se{s_1}, \se{s_2} ) }_{p\in \T}) &= s_1 s_2 \\
        \mathtt{recon}( \tub{ \mathtt{inv} ( \se{s_1} )}_{p\in \T} ) &= \frac{1}{s_1}
    \end{align}
\end{itemize}

The problem of the paper is to define a real number secret sharing scheme which satisfies the listed requirements assuming that each participant follows the protocol.

\section{Proposed Method}\label{sec:proposed}
As mentioned already, we take great inspiration from Shamir's SSS \cite{Shamir}, when proposing our real number SSS.
To give some intuition, we explain the derivation of the proposed scheme in comparison to Shamir's scheme.

The approach in Shamir's scheme is to start by choosing $t$ coefficients $\tub{c_j}_{j \in T}$, where $T = \{ 1,\ldots, t \}$, from \F uniformly and afterwards defining the polynomial
\begin{equation}
    f_s(x) = \br{s + \sum_{j\in T} c_j x^j} \mod q,
\end{equation}
where $s \in \F$ as usual is the secret. 
The shares of $s$ are then defined as
\begin{equation}
\SE{s}= \tub{f(p)}_{p\in \p} .
\end{equation}

For the real number SSS we want to avoid modular arithmetic and have $s\in \R$ and therefore one idea is to write each share $\se{s}$ as
\begin{equation}\label{eq:shamir1}
    \se{s} = s + \sum_{j \in T} c_j p^j,
\end{equation}
where each $c_j$ is Gaussian distributed. We choose the Gaussian distribution because this is the maximum entropy distribution for a random variable on the real number line having a finite mean and variance, \cite[p. 413]{Cover}.

\begin{figure*}[t!]
    \centering
    \begin{subfigure}[t]{0.75\textwidth} 
        \centering
        \input{plotG1}
        \caption{$t=5$.}
    \end{subfigure}
    \begin{subfigure}[t]{0.75\textwidth} 
        \centering
        \input{plotG4}
        \caption{$t=10$.}
    \end{subfigure}
    
    \caption{$n=11$ shares of the secret $s=5.0$ with the threshold $t=5$ for (a) and $t=10$ for (b), where $\se{s} = f(i)$, with $f(x)$ being a polynomial with $t$ coefficients normally distributed with mean value zero and variance 100. }
    \label{fig:G5}
\end{figure*}
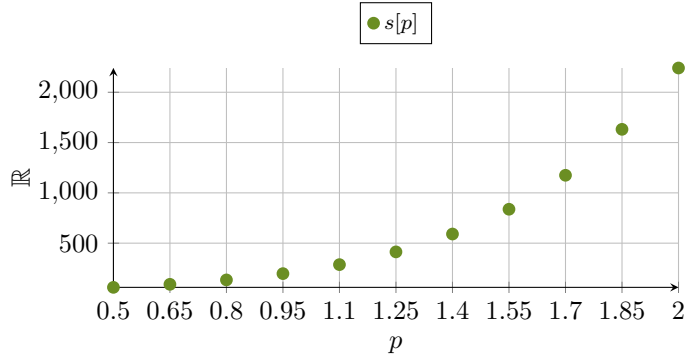
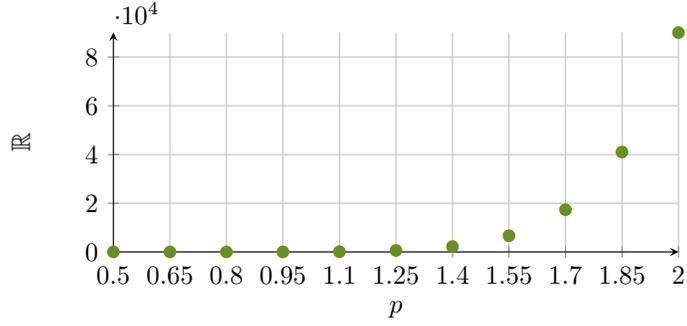

\figref{fig:G5} depicts the shares of a secret $s=5.0$ for $n=11$ participants with $t=5$ (and for comparison also $t=10$). For the Gaussian distributed coefficients, we use mean value zero and variance 100. As seen, the shares seem quite systematic which is not advantageous from a privacy point of view. Specifically, as seen in \eqref{eq:shamir1} the random numbers (the coefficients) are scaled according to $p \in \p$. Consequently, less weight are given to the random numbers of the shares constructed with the lower $p$ values. Therefore, the shares tend to be in numerical order as observed in \figref{fig:G5}.

We can information theoretically verify that the information leak caused by a share decreases as the numerical value of $p \in \p$ increases. Consider for instance $\p = \{1,2,3\}$ and $t=2$, then according to \eqref{eq:shamir1}, the shares of $s$ are
\begin{equation}
    \begin{aligned}
        s[1] &= s + c_1 + c_2 \\
        s[2] &= s + 2 c_1 + 4 c_2 \\
        s[3] &= s + 3 c_1 + 9 c_2.
    \end{aligned}
\end{equation}
Then, assuming $s, c_1,$ and $c_2$ are independent and Gaussian distributed with mean value zero and variance $\sigma^2_s, \sigma^2_{c_1},$ and $\sigma^2_{c_2}$, respectively, then
\begin{equation}\label{eq:mutK}
    \begin{aligned}
        I(S, S[1]) &= \frac{1}{2} \log \left(1+\frac{\sigma_2^{S}}{\sigma^2_{c_1}+\sigma^2_{c_2}} \right) \\
        I(S, S[2]) &= \frac{1}{2} \log \left(1+\frac{\sigma_2^{S}}{4\sigma^2_{c_1}+16\sigma^2_{c_2}} \right)\\
        I(S, S[3]) &= \frac{1}{2} \log \left(1+\frac{\sigma_2^{S}}{9\sigma^2_{c_1}+81\sigma^2_{c_2}} \right).
    \end{aligned}
\end{equation}
\eqref{eq:mutK} clearly shows that the mutual information, and hence, information leakage about the secret, decreases as $p$ increases. 
This does not happen in Shamir's SSS because of the modular arithmetic. We therefore need to adjust the method for it to work in a real number SSS.

To make sure each random number carry the same weight across shares, we propose to construct the shares based on Lagrange interpolation \cite{cd} and we briefly state this method in our notation.

Consider the points $(\alpha_1,\beta_1),\ldots,(\alpha_{t},\beta_{t})$ on the plane $\mathbb
{R}^2$. A polynomial $f(x)$ of at most degree $t-1$, that passes through the points, can be found by
\begin{equation}
    f(x) = \sum_{j\in T} \beta_j L_j(x),
\end{equation}
where $T = \tub{1,\ldots,t}$ and $L_k(x)$ are Lagrange basis polynomials given by
\begin{equation}
    L_j(x) = \prod_{k\in T \backslash \{j\}} \frac{x-\alpha_k}{\alpha_j- \alpha_k}.
\end{equation}

To create shares of a secret, we choose $t$ shares at random and interpolate these shares to a degree (at most) $t$ polynomial $f_s(x)$, by also using that $f_s(0) = s$. Using Lagrange basis polynomials stated above, $f_s(x)$ is written as

\begin{equation}\label{eq:fs}
\begin{aligned}
    f_s(x) &= s \underbrace{ \prod_{k=1}^{t} \frac{x-x_k}{x_0-x_k}}_{L_0(x)} + y_1 \prod_{k=0, k\neq 1}^{t} \frac{x-x_k}{x_1-x_k} 
    + \cdots + y_t \prod_{k=0}^{t-1} \frac{x-x_k}{x_t-x_k} \\
    &= s L_0(x) + y_1 \frac{x}{x_1} \prod_{k=2}^{t} \frac{x-x_k}{x_1-x_k} 
    + \cdots + y_t \frac{x}{x_t}  \prod_{k=1}^{t-1} \frac{x-x_k}{x_t-x_k} \\
    &= s L_0(x) + y_1 \frac{x}{x_1} L_1(x) + \cdots + y_t \frac{x}{x_t}  L_t(x) \\
    &= s L_0(x) + \sum_{j \in T} y_j \frac{x}{x_j} L_j(x),
    \end{aligned}
\end{equation}
where we use that $x_0 = 0$.
The shares are then defined as $$\SE{s} = \{f_s(p)\}_{p\in \p}.$$

\begin{figure*}[t!]
    \centering
    \begin{subfigure}[t]{0.75\textwidth}
        \centering
        \input{plotG2}
        \caption{$t=5$.}
        \label{fig:G2}
    \end{subfigure}
    \begin{subfigure}[t]{0.75\textwidth} 
        \centering
        \input{plotG3}
        \caption{$t=10$.}
        \label{fig:G3}
    \end{subfigure}
    
    \caption{$n=11$ shares of the secret $s=5$ with the threshold $t=5$ for (a) and $t=10$ for (b), where $\se{s} = f(i)$, with $f(x)$ being a polynomial. $t$ points of $f(x)$ (marked with a $ \color{red}{\square}$) are normally distributed with mean value zero and variance $100$.}
    \label{fig:G4}
\end{figure*}
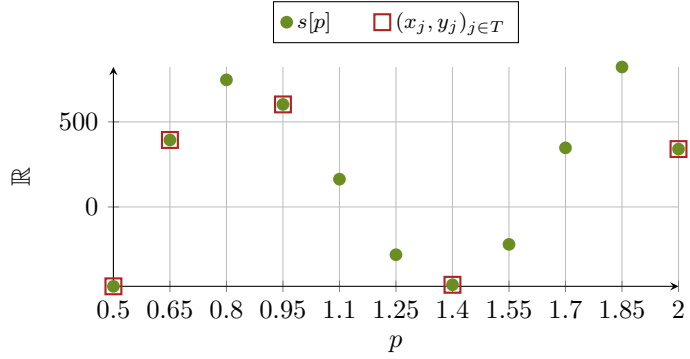
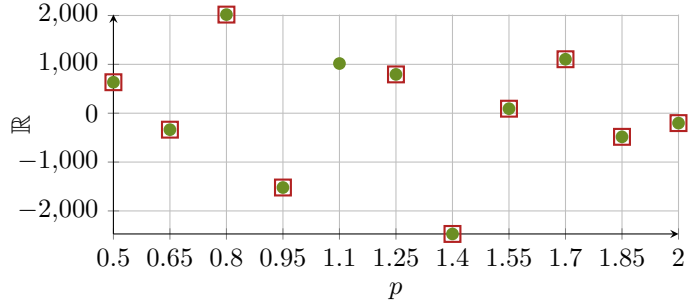

As seen in \eqref{eq:fs}, the random numbers ($y_j$) are normalized and thus have the same weight across shares. Therefore, the shares are much less predictable (especially as $t$ increase) as observed in \figref{fig:G4}.

Moreover, we can information theoretically show that the information leakage of the shares constructed by \eqref{eq:fs} does not depend on the numerical value of $p$. Consider, $\p = \{1,2,3 \}$, $t=2$ and $x_1 = 1$ and $x_2 = 3$. Remark that $x_j$ are chosen each time shares of a secret are constructed and their value is unknown to any adversary. According to \eqref{eq:fs}, the shares of $s$ are written as
\begin{equation}\label{eq:shares0}
    \begin{aligned}
        s[1] &= s \frac{1-1}{0-1} \frac{1-3}{0-3}+y_1\frac{1-0}{1-0} \frac{1-3}{1-3}+y_2\frac{1-0}{3-0}\frac{1-1}{3-1} = y_1\\
        s[2] &= s \frac{2-1}{0-1} \frac{2-3}{0-3}+y_1\frac{2-0}{1-0} \frac{2-3}{1-3}+y_2\frac{2-0}{3-0}\frac{2-1}{3-1} 
        = \frac{1}{3}s-y_1+\frac{1}{3}y_2 \\
        s[3] &= s \frac{3-1}{0-1} \frac{3-3}{0-3}+y_1\frac{3-0}{1-0} \frac{3-3}{1-3}+y_2\frac{3-0}{3-0}\frac{3-1}{3-1} = y_2.\\
    \end{aligned}
\end{equation}
Assuming $s$ and $y_i$ are independent and Gaussian distributed with mean zero and variance $\sigma^2_s, \sigma^2_{Y_1}$, and $\sigma^2_{Y_2}$, respectively, the mutual information yields

\begin{equation}\label{eq:mutK1}
    \begin{aligned}
        I(S, S[1]) &= 0 \\
        I(S, S[2]) &= \frac{1}{2} \log(1+\frac{\frac{1}{9}\sigma^2_{S}}{\sigma^2_{Y_1}+\frac{1}{9}\sigma^2_{Y_2}})\\
        I(S, S[3]) &= 0.
    \end{aligned}
\end{equation}

Thus, \eqref{eq:mutK1} shows that the information leakage caused by the shares are independent of the numerical value of $p$, which is of course important from a privacy perspective. More precisely, the difference between \eqref{eq:mutK} and \eqref{eq:mutK1}, is that in the former each participant $p$ knows that you gain most information about the secrets the lower the value of $p$ you have. In the latter, it is unknown to the participants which shares have zero mutual information and which does not, and it is different for each secret.

We state the $\mathtt{share}$ algorithm of the real number SSS formally in \proref{pro:Secret} and expand on the privacy analysis of it in section \ref{sec:privacy_analysis}.

\begin{pro}{H}
	\caption{$\mathtt{share}(s,t,\p) = \SE{s}$}
	\label{pro:Secret}
		\begin{algorithmic}[1]
		\Statex Input: $s$ is the secret, $t$ is the threshold and $\p$, with $|\p| = n$ is the index set of the participants.
		
		\Statex Output: $\SE{s}$ is the set of shares of $s$.
		
		\State Draw distinct $\{x_j\}_{j\in T}$ from $\p$, where \m{T = \tub{1,\ldots, t}.}
		
        \State $y_j \overset{\text{iid}}{\sim} \N(\mu_Y, \sigma_Y^2)$ for $j \in T $, where $\mu_Y$ and $\sigma_Y^2$ are chosen privacy parameters.
        
        \State $f_s(x) = s L_0 + \sum_{j \in T} y_j \frac{x}{x_j} L_j(x)$.
        
        \State $\SE{s} = \{f_s(p)\}_{p\in \p}$.
		
		\end{algorithmic}
\end{pro}
Remark, that \proref{pro:Secret} has two privacy parameters $\mu_Y$ and $\sigma^2_Y$ which can be chosen by the party constructing shares of its secret. The mean value does not have a significant effect on the privacy and could in principle be chosen randomly (or as zero as we do throughout the paper). In section \ref{sec:privacy_analysis} the impact of $\sigma^2_Y$ becomes clear.

The reconstruct algorithm of the proposed real number secret sharing scheme, is almost identical to the one of Shamir's scheme (the only difference is the lacking of modular arithmetic). Since the algorithm consist solely of Lagrange interpolation, we state it without further introduction.
\begin{pro}{H}
	\caption{$\mathtt{recon}(\TE{s}{p}) = \hat{s}$}
	\label{pro:recon}
		\begin{algorithmic}[1]
		\Statex Input: $\TE{s}{p}$, with $|\T| > t$, is a set of at least $t+1$ shares of s.
		
		\Statex Output: $\hat{s}$, the reconstructed secret.
        
        \State Define $\{(p, \se{s})\}_{p\in \T}$ as the set of points to interpolate. 
        
        \State $f_r(x) = \sum_{p \in \T} s[p] L_p(x)$.
        
        \State $\hat{s} = f_r(0)$.
		
		\end{algorithmic}
\end{pro}

To be clear, our proposed real number secret sharing scheme, consists of the algorithms $\mathtt{share}$ and $\mathtt{recon}$ stated in \proref{pro:Secret} and \proref{pro:recon}, respectively. To give intuition about the proposed method, Example \ref{exa:1} gives an example of using it. 
\begin{exa}[Real number secret sharing]\label{exa:1}
Let $s = 5.0$ be a secret and $ \scriptstyle \p = \tub{0.5,  0.65, 0.8,  0.95, 1.1,  1.25, 1.4,  1.55, 1.7,  1.85, 2}$ the index of the participants. Consider $\mathtt{share}(s,t,\p)$ in \proref{pro:Secret} to create shares of $s$ for $n=11$ participants. We perform the following steps with $t = 5$:
\definecolor{red}{RGB}{178,34,34}
\definecolor{dgreen}{RGB}{107,142,35}
\begin{enumerate}
    \item $\tub{x_j}_{j\in T} = \tub{0.5, 0.65, 0.95, 1.4, 2}$.
    \item $\tub{y_j}_{j\in T} = \scriptstyle \tub{-466.506, 393.646, 602.653, -457.489, 340.160}$. See $(x_j, y_j)_{j \in T}$ in \figref{fig:G2} marked with $ \color{red}{\square}$. 
    \item Define $f(x) = s L_0 + \sum_{j \in T} y_j \frac{x}{x_j} L_j(x)$.
    \item Define $\{\se{s}\}_{p\in \p} = \{f(p)\}_{p\in \p}$. See $\SE{s}$ in \figref{fig:G2} marked with \tikz\draw[dgreen,fill=dgreen] (0,0) circle (.5ex);. 
\end{enumerate}
For comparison, we perform the same steps for $t=10$, where $(x_j, y_j)_{j \in T}$ are seen in \figref{fig:G3} marked with $ \color{red}{\square}$ and $\SE{s}$ are seen in \figref{fig:G3} marked with \tikz\draw[dgreen,fill=dgreen] (0,0) circle (.5ex);.

\end{exa}

We will now show that the scheme satisfies the requirements listed in section \ref{sec:problem}. We start by noting that the proof of Lagrange interpolation also proves the correctness of the scheme. Therefore, we immediately analyse the privacy of the scheme in the following section.

\subsection{Privacy Analysis} \label{sec:privacy_analysis}

We start out the analysis by considering one participant $p\in \p$, who does not know $s$, but learns $\se{s}$. That is, from the view of $p$, $s$ can be modeled as the outcome of the random variable $S$ having some distribution. The uncertainty $p$ has about $s$ can be stated as the differential entropy $h(S)$ of $S$. Also $\se{s}$ is the outcome of a random variable $\se{S}$. To see the relation between $S$ and $\se{S}$, consider the rewrite of $\se{s}$ 
\begin{equation}
\begin{aligned}
    \se{s} &= s L_0(p) + \underbrace{\sum_{j \in T} y_j \frac{p}{x_j} L_j(p)}_{b(p)} 
    = s L_0(p) + b(p),
\end{aligned}
\end{equation}
To this end, we have that
\begin{equation}\label{eq:s(i)}
    \se{S} = S L_0(p) + B(p).
\end{equation} 
We choose to model the $L_j$ values as constants even though it can be argued that they are indeed random variables because each $x_k$ from step 1. of \proref{pro:Secret} are randomly chosen. However, since $t$ is generally close to $n$ and $\p$ is public, there is not an insignificant probability of guessing the $x_k$ values. Consider for instance $\p$ given in Example \ref{exa:1} and let $t = 10$. Then we know that $x_{1} \in \{0.5, 0.65\}$ because the 9 remaining $x_k$ values must also be distinct elements of $\mathcal{P}$. Consequently, for each $x_k$ there are generally only a few possible values it can take and thus in our analysis we choose to treat each $L_j$ value as a constant.
To this end, $B(p)$ is normally distributed
with mean $t \mu_Y$ and variance 
\begin{equation}
    \sigma_{B(p)}^2 = \sigma_Y^2 \sum_{j\in T} \br{\frac{p}{x_j} L_j(p)}^2 .
\end{equation} 

Consider now the mutual information $I(S; \se{S})$ between $S$ and $\se{S}$;
\begin{equation}\label{eq:entr}
\begin{aligned}
    I(S;\se{S}) &= h(\se{S}) - h(\se{S}|S) \\
    &= h(SL_0(p) +B(p)) - h(SL_0(p)+B(p) | S) \\
    &= h(SL_0(p) +B(p)) - h(B(p)),
\end{aligned}
\end{equation}
where we use that $I(X,Y)$ is symmetric, that $L_0$ is a constant, and that $S$ and $B(p)$ are independent. Before we proceed, we remark that when $f_s(x)$ given in step 3. of \proref{pro:Secret}, is evaluated in one of the $x_k$ values chosen in step 1., $y_k$ is outputted. That is, $f_s(x_k) = y_k$, see \eqref{eq:shares0}. Recall that each $y_k$ is Gaussian distributed and since $x_k \in \p$, we have that exactly $t$ shares are completely independent of the secret $s$. Thus, in this best case scenario, which is true for $t$ shares, $I(S; \se{S}) = 0$ and there is no leak of information.
To analyse the information leakage of the remaining $n-t$ shares, we take the same approach as in \cite{jane} and consider again \eqref{eq:entr}. As discussed, $B(p)$ are Gaussian distributed and according to \cite[p. 244]{Cover}, the differential entropy of a Gaussian distributed variable $X$ can be written as
\begin{equation}
    h(X) = \frac{1}{2} \log_2(2\pi e \sigma^2_X),
\end{equation}
where $e$ is the Euler number. On the other hand, we do not make an assumption of the distribution of $\m{SL_0(p) + B(p)}$, since this can vary from application to application. Instead, we note that a high entropy of $SL_0(p) + B(p)$, results in a higher $I(S;\se{S})$ in equation \eqref{eq:entr}. Thus, by using the maximum entropy distribution (which is the Gaussian distribution) as the distribution of $SL_0(p) + B(p)$, we establish an upper bound on the mutual information.
\begin{equation} \label{eq:mutual}
\begin{aligned}
        I(S;\se{S}) &= h(SL_0(p) +B(p)) - h(B(p)) \\
        & \leq \frac{1}{2} \log(2\pi e (\sigma_{S_{L_0(p)}}^2 + \sigma_{B(p)}^2 )) - \frac{1}{2} \log(2\pi e \sigma_{B(p)}^2) \\
        &= \frac{1}{2} \log\left(1 + \frac{\sigma_{SL_0(p)}^2} {\sigma_{B(p)}^2}\right),
\end{aligned}
\end{equation}
where we use that since $S$ and $B(p)$ are independent, the variance of $\se{S}$ can be written as 
\begin{equation}
    \sigma_{\se{S}}^2 = \sigma_{SL_0(p)}^2 + \sigma_{B(p)}^2.
\end{equation}

In conclusion, choosing for instance $\sigma^2_{B(p)}$ 100 times larger than the variance of $\sigma^2_{SL_0(p)}$, the leaked information is at most $0.0072$ bits (no matter the real distribution of $SL_0(p) + B(p)$), which is to be read in the way that \textit{on the average} one share of $s$ leaks $0.0072$ bits. For comparison, if the secret indeed is Gaussian distributed with variance 10, the uncertainty about it is $3.7080$ bits and after learning $\se{s}$, the uncertainty is $3.7008$ bits. Hence, each share $\se{s}$ leaks only very little information about $s$, when choosing the variance $\sigma^2_Y$ large enough. 

To continue this analysis, note that in the problem statement, we require that a set of at most $t$ shares should reveal very little information about the secret. Thus, we now address the mutual information between $s$ and a set of $t$ shares. That is,
\begin{equation}\label{eq:mut}
\begin{aligned}
    I(S; S[1],\ldots, S[t]) &= h(S[1],\ldots,S[t])
    - h(B(1),\ldots,B(t))
\end{aligned}
\end{equation}
Again, we notice that in the best case scenario, the set of $t$ shares is exactly the set of normally distributed values $y_j$ chosen in step 2. of \proref{pro:Secret}, i.e. $\TE[\T']{s}{p} = \{y_j\}_{j\in T}$, where $T={1,\ldots, t}$. In this case, all $t$ shares are independent of the secret $s$ and thus we have no leak of information. This case happens with a high probability if $t$ is close to $n$. However, due to properties of the scheme, which we will explore in the following section, $t$ might be chosen less than $\lfloor \frac{n}{2} \rfloor$. In this case, we may have that none of the $t$ shares are independent of the secret.  
This would be the worst case scenario, which we address now by establishing an upper bound for $I(S; S[1],\ldots, S[t])$ by using the same trick as previously. Namely, we choose the $t$-variate Gaussian distribution for $X_S = (S[1],\ldots, S[t])$, which is the maximum entropy distribution. Since the sum of two Gaussian distributions is still Gaussian, we have that $X_B = (B(1),\ldots,B(t))$ also follows a $t$-variate Gaussian distribution. The entropy of a $N$-variate Gaussian distributed variable $X$ is given as \cite[p.249]{Cover}
\begin{equation}
     h(X) = \frac{1}{2} \log\left(( 2\pi e)^N \det(C_X)\right), 
\end{equation}
where $C_X$ is the covariance matrix for $X$.
This expression can be used directly in \ref{eq:mut}, yielding
\begin{equation}\label{eq:mut1}
\begin{aligned}
    I(S; X_S) &\leq \frac{1}{2} \log\left((2 \pi e)^t \text{det}(C_{X_S})\right) 
    - \frac{1}{2} \log\left((2 \pi e)^t \det(C_{X_B}) \right) \\
    &= \frac{1}{2} \log\br{(2 \pi e)^t \frac{\det(C_{X_S})}{\det(C_{X_B})}} 
\end{aligned}
\end{equation}

Using \eqref{eq:s(i)}, we can write the $(i,j)$'th term of the covariance matrix $C_{X_S}$, as
\begin{equation}
\begin{aligned}
    c_{X_S}(i,j) &= \cov{S{L_0(i)} + B(i) ; S{L_0(j)} + B(j)  } \\
    &= \cov{S{L_0(i)} ; S{L_0(j)} + B(j) } 
    + \cov{B(i) ; S{L_0(j)} + B(j)  } \\
    &= \cov{S{L_0(i)} ; S{L_0(j)} } + \cov{S{L_0(i)} ; B(j)  } \\
    &+ \cov{B(i);S{L_0(j)}   } + \cov{B(i) ; B(j)  } \\
    &= \cov{SL_0(i); SL_0(j)} + \cov{B(i) ; B(j)  },
\end{aligned}
\end{equation}
where we use that $SL_0(i)$ and $B(i)$ are independent. Therefore, we have
\begin{equation} \label{Eq.SumCovariance}
C_{X_S} = C_{X_{SL_0}} + C_{X_B},    
\end{equation}
where $X_{SL_0} = (SL_0(1),\ldots, SL_0(t))$.

Thus, analogue to the previous result, the leaked information is controlled by the relation between the variance of $S$ and the variance of $Y$. By choosing $\sigma_{Y}^2$ large compared to the variance of $S$, the determinant of $C_{X_S}$ will be only slightly larger than the determinant of $C_{X_B}$ and we have that asymptotically, the leaked information goes to zero bits. 

We can therefore make the following proposition, stating that the scheme fulfills the privacy requirement.
\begin{prop}
The real number secret sharing scheme comprised of the algorithms $\mathtt{share}$ and $\mathtt{recon}$ stated in \proref{pro:Secret} and \proref{pro:recon}, respectively, satisfy that for any $\delta >0$ there exists the covariance matrix $C_{X_B}$ such that  
 \begin{equation}
        I(S; \tub{\se{S}}_{i \in \T'}) \leq \delta,
    \end{equation}
for a secret $s$ being the outcome a random variable $S$ and shares $\TE[\T']{s}{p}$ being the outcome of the random variables $\TE[\T']{S}{p}$.  
\end{prop}
\begin{proof}
We use \eqref{eq:mut1} and \eqref{Eq.SumCovariance}. For short, we write ${A = C_{X_{SL_0}} }$, and $B = C_{X_B}$. Since $A$ and $B$ are symmetric, in fact positive semi-definite, they can be simultaneously diagonalizable. We denote the eigenvalues of $A$ by $\lambda_A^i$, and $B$ by $\lambda_B^i$. Let $\overline \lambda_A, \overline \lambda_B$ be the maximal eigenvalue of $A,B$ respectively, and $\underline \lambda_B$ be the minimal eigenvalue of $B$. 
Specifically,
\begin{align*}
 \frac{\det(C_{X_S})}{\det(C_{X_B})}  =   \frac{\prod_{i=1}^N (\lambda_A^i + \lambda_B^i)}{\prod_{i=1}^N \lambda_B^i } \leq \frac{\overline \lambda_A + \overline \lambda_B}{\underline \lambda_B} = \frac{\overline \lambda_A}{\underline \lambda_B} + \frac{\overline \lambda_B}{\underline \lambda_B}.
\end{align*}

Hence, by rescaling $\det(C_{X_B})$ by sufficiently large coefficient, the mutual information $I(S; S(1),\ldots, S(t))$ can be made arbitrarily small.
\end{proof}

In the next section, we will show that the real number secret sharing scheme also satisfies that final requirement. 

\subsection{Computations on Shares}
In this section, we will define the algorithms $\mathtt{add}$, $\mathtt{mult}$, and $\mathtt{inv}$, which perform addition, multiplication and inverse of secrets directly on the shares and, thus, does not leak any secrets. To improve readability and intuition, we present the operations using scalars, however the methods are easily extendable to matrices as well. To this end, we start by defining what we mean by shares of a matrix (and equivalent; a vector).
\begin{defi}[Secret shared matrix]
Let $\v A \in \R^{m_1\times m_2}$ be a matrix and let each entry of \v A be secret shared using $\mathtt{share}$. To this end, $\se{\v A}$ denotes the matrix consisting of the $p$'th share of each element in $\v A$, respectively.
\end{defi}

For the rest of this section, assume that $s,a \in\R$ are secrets and that each participant $p \in \p$ holds the shares $\se{s}$ and $\se{a}$, respectively.

\subsubsection{Addition}
We start out with the simplest operation, which is addition. The output of the addition algorithm is that each participant $p$ holds a share $\se{c}$, where $c = s+a$. Note that since each share is a point on a polynomial, it can be written as
\begin{equation}
    \se{s} = s + \alpha_1 p + \alpha_2p^2 +\cdots+\alpha_t p^t,
\end{equation}
and 
\begin{equation}
    \se{a} = a + \beta_1 p + \beta_2p^2 +\cdots+\beta_t p^t,
\end{equation}
where $\alpha_j, \beta_j \in  \R$ are coefficients. Adding the above expressions yields
\begin{equation}
   \se{c} = \se{s} + \se{a} = (s + a) + (\alpha_1+\beta_1) p +\cdots+(\alpha_t+\beta_t) p^t.
\end{equation}

Hence, by participant $p$ performing $\se{s} + \se{a}$, it now holds a share $\se{c}$, where $\m{c=s+a}$. To denote the computation of adding shares we simply use the '$+$' sign or we write $\m{\mathtt{add}( \se{s}, \se{a}) = \se{c}}$. Note that subtraction is performed on the shares equivalently, which we simply denote by '$-$'. 

\subsubsection{Multiplication}
Multiplying shares is somewhat more complicated. If we attempted to simply multiply the polynomials like we added them previously, we would find that the degree of the resulting polynomial is $2t$. In this case we need $\m{2t +1}$ shares to reconstruct the secret. To avoid the growing degree of the polynomial, we use a well-know trick called Beavers' trick, \cite{beaver}. It uses so-called triplets, $\{\se{r_1}, \se{r_2}, \se{r_1r_2}\}_{p\in \p}$ of shares of (unknown) random numbers $r_1$ and $r_2$, and their product $r_1r_2$. To create the triplets, it is typically required that $t < \lfloor \frac{n}{2} \rfloor$, however there are ongoing research in efficient methods of generating Beaver triplets for larger values of $t$, \cite{imprSPDZ}.

We state formally $\mathtt{mult}$ in \proref{pro:mult}.

\begin{pro}{H}
	\caption{$\mathtt{mult}( \se{s}, \se{a} ) = \se{sa}$ }
	\label{pro:mult}
		\begin{algorithmic}[1]
		\Statex Input: $\SE{s}$,$\SE{a}$  shares of the secrets and $\m{\{ \se{r_1}, \se{r_2}, \se{r_1r_2}\}_{p\in \p}}$ shares of the unknown Beavers triplet.
		
		\Statex Output: $\SE{sa}$, shares of the product of the secrets.
		
		\State 
		\begin{align}
           d &= \mathtt{recon}( \tub{  \se{s}  - \se{r_1}  }_{p\in \T} ) \\
    e &= \mathtt{recon}( \tub{ \se{a}  -\se{r_2} }_{p\in \T} ) ,
        \end{align}
        
        \State 
        \begin{equation}
            \se{sa} = de + d \se{r_2}  + \se{r_1} e + \se{r_1r_2},
        \end{equation}
        
		\end{algorithmic}
\end{pro}

To see that the multiplication protocol in \proref{pro:mult} is correct, perform the following rewrite
\begin{equation}
    \begin{aligned}
        s &= d + r_1  \\
        a &= e + r_2,
    \end{aligned}
\end{equation}
to see that 
\begin{equation}
        sa[p] = (d+r_1[p])(e+r_2[p]).
\end{equation}
Note that a public constant (like $e$ and $d$ in this case) can be directly multiplied on the shares by each participant. This can easily be verified by using the same approach as showing that the $\mathtt{add}$ protocol is correct.

We also remark that $d = s + r_1$ and $e = a + r_2$ are revealed in plain text in \proref{pro:mult}. Since $r_1$ and $r_2$ are Gaussian distributed, $d$ and $e$ does not leak more than a share of the secrets. However, we give here the upper bound of the information leak of knowing both $t$ shares of $s$ and also $d$.  
\begin{equation}\label{eq:MImult}
\begin{aligned}
    I(S&; \TE[\T']{S}{p}, S+R_1)\\
    &= h(\TE[\T']{S}{p},S+R_1) - h(\TE[\T']{S}{p},S+R_1|S) \\
    &= h(\TE[\T']{S}{p},S+R_1) \\
    &- h(\tub{SL_0(p) + B(p)}_{p\in \T'},S+R_1|S)\\
    &= h(\TE[\T']{S}{p},S+R_1) - h(\tub{B(p)}_{p\in \T'},R_1)
\end{aligned}
\end{equation}
To find an upper bound on the information leakage we use the maximal entropy distribution for the distribution of ${X_{SR_1} = (\TE[\T']{S}{p},S+R_1)}$. By design, $\m{X_{BR_1} = (\tub{B(p)}_{p\in \p},R_1)}$ are distributed according to a multivariate Gaussian distribution. To this end we have,
\begin{equation}\label{eq_ka1}
\begin{aligned}
    I(S; X_{SR_1}) &\leq \frac{1}{2} \log\left((2 \pi e)^t  \frac{\det(C_{X_{SR_1}})} { \det(C_{
    X_{BR_1}})} \right),
    \end{aligned}
\end{equation}
where $C_{X_{SR_1}}, C_{X_{BR_1}}$ are the covariance matrices of $X_{SR_1}$ and $X_{BR_1}$, respectively. As seen, the result in \eqref{eq_ka1} is very similar to the one obtained in \eqref{eq:mut1}.
To demonstrate the (at most) revealed data using $\mathtt{mult}$, Example \ref{exa:mult} demonstrates the multiplication of two secrets. Note that in section \ref{sec:numerical_eval} we numerically estimate the leak of information caused by the multiplication protocol.
\begin{exa}[Multiplication of shares]\label{exa:mult}
Let the number of participants $n = 7$, $\p = \{1,2,\ldots, 7 \}$, and the threshold $t = 3$. Consider two secret $s_1 = 34.5$ and $s_2 = 3.42$ and the multiplication of them performed on their shares. To demonstrate the (small) information leak caused by $\mathtt{mult}$, \figref{fig:examult} depicts $t =3$ shares of each secret and the values $d$ and $e$ revealed by the algorithm. As seen, it is very hard to deduce the true values of the secrets using the revealed information. 
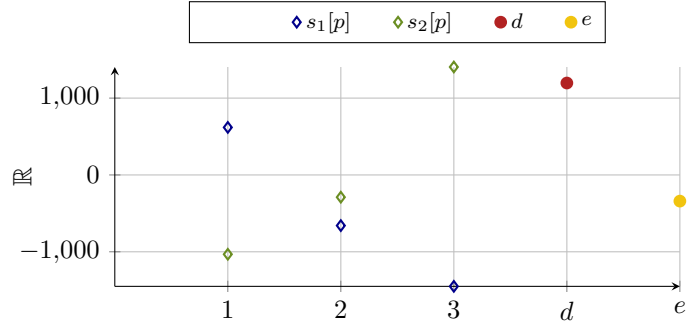
\begin{figure}
    \centering
    \input{plotExaMult}
    \caption{Example of the information known about two secrets $s_1$ and $s_2$ after executing the $\mathtt{mult}$ algorithm. In the worst case, the adversary knows $t = 3$ shares of each of the secrets and the values $d$ and $e$ revealed by $\mathtt{mult}$. In this example $s_1 = 34.5$ and $s_2 = 3.42$, which is very hard to deduce from the revealed information.}
    \label{fig:examult}
\end{figure}
\end{exa}

We use $\mathtt{mult}( \se{s}, \se{a}) = \se{sa})$ to denote the computation of multiplying shares using Beaver's trick. In continuation, we note that $\mathtt{mult}$ can easily take two matrices as input, for instance $\se{\v A}$ with $\v A \in \R^{m_1 \times m_2}$ and $\se{\v B}$ with $\v B \in \R^{m_2 \times m_3}$. In this case the Beavers triplet is also matrices; $\v R_1 \in \R^{m_1 \times m_2}$, $\v R_2 \in \R^{m_2 \times m_3}$ and $\v R_1 \v R_2$ is the matrix-matrix product. The rest of algorithm \ref{pro:mult} remains unchanged.

\subsubsection{Division}
We consider the inversion $s^{-1}$ and note that one could afterwards use $\mathtt{mult}$ to compute a secret divided by another secret. We propose to compute this operation efficiently on the shares, by noting that
\begin{equation}\label{eq:39}
     s^{-1} = \frac{1}{sr} r,
\end{equation}
where $r \in \R$ is a random number. To this end, we propose to use a normally distributed random variable $r$ which is unknown to the participants. This $r$ can be constructed in the following way; each participant $p$ chooses a Gaussian distributed value $r_p$ and distributes the shares $r_p[j]$ to participant $j \in \p$. Each participant $p$ then computes its share of $r$ by $\se{r} = \sum_{j \in \p} r_j[p]$. 

To calculate \eqref{eq:39}, the participants use \m{\mathtt{recon}(\mathtt{mult}(\se{s} , \se{r})) = sr} to learn in plain text the product $sr$. Subsequently, they each compute $s^{-1}[p] = \frac{1}{sr} \se{r}$ to learn individual shares of $s^{-1}$. To improve readability, we state the division algorithm in \proref{pro:div}.
\begin{pro}{H}
	\caption{$\mathtt{inv}(\se{s}) = s^{-1}[p]$ }
	\label{pro:div}
		\begin{algorithmic}[1]
		\Statex Input: $\SE{s}$ shares of the secret and $\SE{r}$ shares of an unknown random value $r\in \R$.
		
		\Statex Output: $\SE{s^{-1}}$, shares of the inverse secret.
        
        \State $sr = \mathtt{recon}(\mathtt{mult}(\se{s}, \se{r} )$. 
        
        \State $\se{s^{-1}} = (sr)^{-1} \se{r}$.
		
		\end{algorithmic}
\end{pro}
We remark that the plain text $sr$ does reveal some information about $s$. However, this information leak can be upper bounded. We here compute the maximal information leak about $s$ from a set of $t$ shares of $s$ joint with $sr$.  
\begin{equation}
\begin{aligned}
    I(S; \TE[\T']{S}{p}, SR)
    &= h(\TE[\T']{S}{p},SR) - h(\TE[\T']{S}{p},SR|S) \\
    &= h(\TE[\T']{S}{p},SR)
    - h(\tub{SL_0(p) + B(p)}_{p\in \T'},SR|S)\\
    &= h(\TE[\T']{S}{p},SR) - h(\tub{B(p)}_{p\in \T'},R)
\end{aligned}
\end{equation}
We do not make assumptions on the joint distribution of $X_{SR} = (\SE{S},SR)$, thus we make an upper bound for the mutual information by choosing the maximal entropy distribution. By design, $\m{X_{BR} = (\tub{B(p)}_{p\in \p},R)}$ are distributed according to a multivariate Gaussian distribution. To this end we have,
\begin{equation}
\begin{aligned}
    I(S; X_{SR}) &\leq \frac{1}{2} \log\left((2 \pi e)^t  \frac{\det(C_{X_{SR}})} { \det(C_{
    X_{BR}})} \right),
    \end{aligned}
\end{equation}
which is a very similar result to the one obtained in \eqref{eq:mut1}.

We denote the computation of $\se{s^{-1}}$ as $\mathtt{inv}(\se{s}) = \se{s^{-1}}$ and note that also $\mathtt{inv}$ can take a matrix as input. In this case the random value $r$ is simply a random matrix of suitable dimension and the rest of the algorithm remains the same.  

\section{Numerical Evaluation}\label{sec:numerical_eval}

In this section we evaluate the numerical performance of the proposed real number secret sharing scheme. To this end, we have implemented the scheme on a laptop PC in the programming language Python that uses the IEEE 754 floating point standard. We start by evaluating the accuracy of the scheme in terms of the variance of the Gaussian distributed $y_j$ values in \proref{pro:Secret}. The parameters we have chosen are $n=11$ participants, $t = 5$, and the secrets $s_1 = 5.5$ and $s_2 = 34.7$. We simulate both $\mathtt{recon}$ (\proref{pro:recon}), $\mathtt{add}$, $\mathtt{mult}$ (\proref{pro:mult}), and $\mathtt{inv}$ (\proref{pro:div}), where we start by generating shares of the secrets using \proref{pro:Secret}. Afterwards, we either directly reconstruct the secret using $\mathtt{recon}$ in \proref{pro:recon} or use respectively, $\mathtt{add}$, $\mathtt{mult}$, or $\mathtt{div}$ on the shares before reconstruction. To evaluate the accuracy, we use the root square error (RSE) between the expected result $v$ and the reconstructed result $\hat{v}$, which is defined as 
\begin{equation*}
\text{RSE} =  \sqrt{ (v - \hat{v} )^2}. 
\end{equation*}

\begin{figure}
    \centering
    \input{plotAccuracy}
    \caption{Accuracy of the algorithms $\mathtt{recon}$, $\mathtt{add}$, $\mathtt{mult}$, and $\mathtt{inv}$ in terms of the variance of the Gaussian distributed $y_j$ values in $\mathtt{share}$ (\proref{pro:Secret}). The loss of accuracy is due to numerical errors.  }
    \label{fig:NumAcc}
\end{figure}
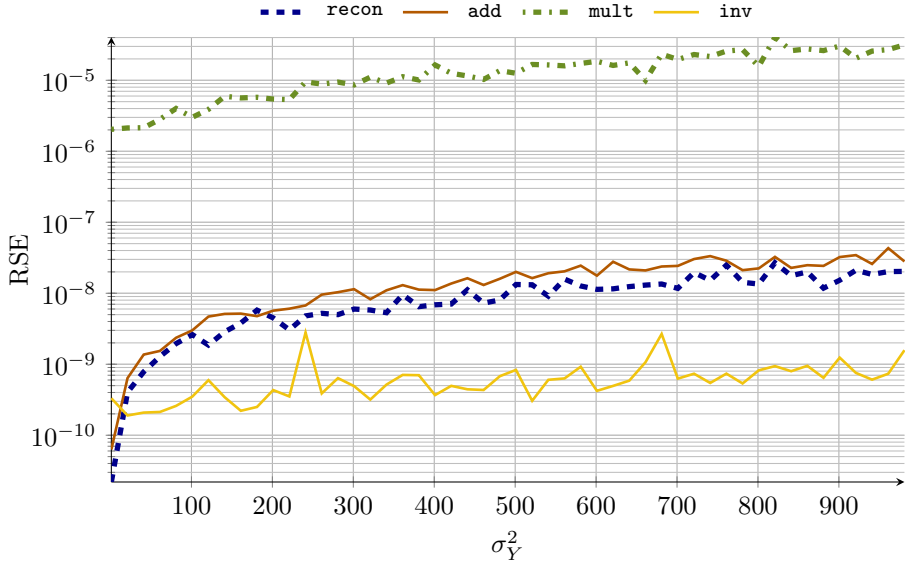

\figref{fig:NumAcc} depicts the RSE between $v$ and $\hat{v}$ as a function of $\sigma^2_Y$, for all four algorithms. As seen, as $\sigma^2_Y$ is increased, the accuracy slowly decreases. This is purely due to numerical errors, because as the $y_j$ values in \proref{pro:Secret} increases, the shares grow exponentially large and consequently loose precision due to the floating point representation. The reason why $\mathtt{inv}$ achieves such high precision, is because the outputted shares are relatively small due to the reciprocal operation of the algorithm.

\begin{figure}
    \centering
    \input{plotMI1share}
    \caption{Mutual information (MI) between a standard normal distributed secret, $S$ and, respectively, one share of $S$, $t$ shares of $S$, and $t$ shares of $S$ joint with $S+R_1$ for a normal distributed variable $R_1$ (see \proref{pro:mult}). }
    \label{fig:NumMI1}
\end{figure}
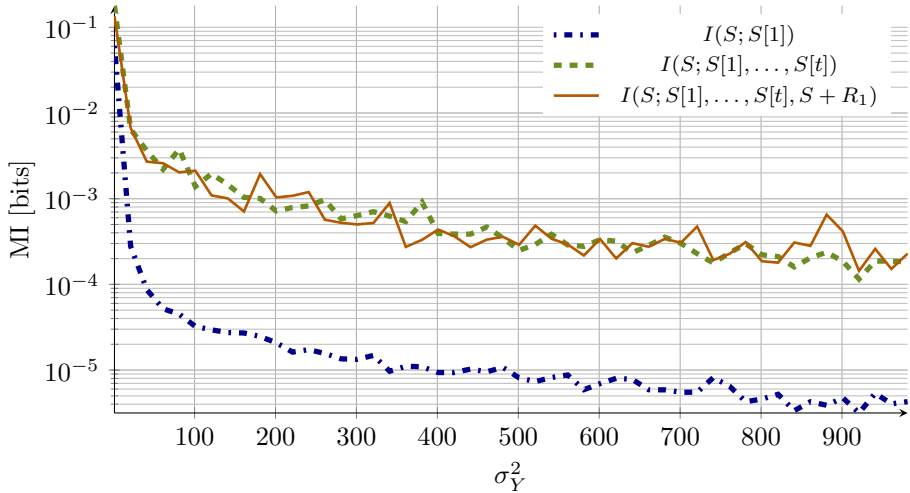

Finally, we numerically evaluate the privacy properties of the scheme. That is, we estimate the privacy loss of the secret from one share, from $t$ shares and from performing multiplication. In particular, we estimate $I(S;S[1])$ in \eqref{eq:mutual}, $I(S; S[1],\ldots,S[t])$ in \eqref{eq:mut} , and $I(S; S[1],\ldots,S[t], S+R_1)$ in \eqref{eq:MImult}, based on simulated data. 
These estimations are a product of statistical analysis, thus we generate a large sample size of each relevant variable for each estimation. We simulate in each case the secret $S \sim \N(0,1)$ and the remaining variables are computed based on the secret. \figref{fig:NumMI1} depicts all three estimations and as expected, one share of the secret leaks very little information while $t$ shares clearly has a greater leak. As seen, these numerical results validate the theoretical results.

\section{Application to Kalman filtering} \label{sec:application}

To demonstrate our proposed privacy preserving computation framework, we use the Kalman filter \cite{KalmanIntro} to estimate $\hat{\v x}_k$ of \eqref{one} when given only real number secret shared versions of the observations in \eqref{two}. The Kalman filter consists of the following 5 equations, where $\v P$ is the covariance matrix of the estimate, \v K is the Kalman gain and $\v Q $ and \v R are covariance matrices of the process and measurement noise respectively, 

\begin{equation}\label{eq:kalman}
\begin{aligned}
    \Tilde{\v x}_k &= \v A \hat{\v x}_{k-1} + \v B \v u_k \\
    \Tilde {\v P}_k &= \v A \v P_{k-1} \v A^{\top} + \v Q_k\\
    \v K_k &= \Tilde{ \v P}_k \v H^{\top} ( \v H_k \Tilde{ \v P}_k \v H^{\top} + \v R_k) ^{-1} \\
    \hat{\v x}_k &= \Tilde{\v x}_k + \v K_k (\v z_k - \v H \Tilde{\v x}_k) \\
    \v P_k &= \Tilde{\v P}_k - \v K_k \v H_k \Tilde{\v P}_k.
   \end{aligned} 
\end{equation}

We consider the following scenario. Assume that $n$ none-colluding entities are used as computing units, hereafter referred to as computing parties. That is, the computing parties perform all computations given only shares of the data. Each time the computing parties receive shares of a new measurement, they compute a new update of the state estimate. We do not specify who delivers these measurements, but it could likely be from a collection of nodes or from a set of other participants. The computing parties are not allowed to learn any clear text data and they only output shares (which can afterwards be reconstruct to the clear text output).

In \proref{pro:RLS}, we state a privacy preserving Kalman filter based on the proposed real number secret sharing scheme, from the view of computing party $p \in \p$.

\begin{pro}{H}
	\caption{$\mathtt{privKalman}() $ }
	\label{pro:RLS} 
		\begin{algorithmic}[1]
		\Statex Input: $\se{\v u_k}$ for all $k$, are shares of the observations, $\v P_0$ and $\v K_0$ can be initialized as identity matrices.
		
		\Statex Output: $\se{\hat{\v x}_k}$; the estimate of the $k$'th state of the system.
		
		\For {all $k$}
		
		\State $ \Tilde{ \v x}_k[p] = \mathtt{mult} ( \v A[p] ,  {\hat{\v x}_{k-1}[p]} ) +  \mathtt{mult} ( \v B[p] ,  \v u_k[p] ) $
		\State $ \v V_k[p] =  \mathtt{mult}( \v P_{k-1}[p], \v A^{\top}[p] )$
		\State $ \Tilde{\v P}_k[p] =  \mathtt{mult} ( \v A[p] , \v V_k[p] ) + \v Q_k[p]$
		\State $\v S_k[p] = \mathtt{mult} ( \v H[p] , \mathtt{mult}( \Tilde{\v P}_k[p], \v H^{\top}[p] ) ) + \v R_k[p]$
		\State $\v K_k[p] = \mathtt{mult} ( \mathtt{mult}( \Tilde{\v P}_k[p], \v H^{\top}[p]), \mathtt{inv}(\v S_k[p]) )$
		\State $\v y_k[p] = \v z_k[p] - \mathtt{mult} ( \v H_k[p], \Tilde{ \v x}_k[p] )$
		\State $ \hat{ \v x}_k[p] = \Tilde{ \v x}_k[p] + \mathtt{mult} ( \v K_k[p], \v y_k[p] ) $
		\State $ \v P_k[p] = \Tilde{ \v P}_k[p] - \mathtt{mult} ( \v K_k[p], \mathtt{mult} ( \v H_k[p], \Tilde{ \v P}_k[p]))$
		\EndFor
		\end{algorithmic}
\end{pro}
Remark that \proref{pro:RLS} does not reveal the result or any intermediate results. 

\subsection{Simulation}

We have simulated \proref{pro:RLS} and compared its estimation performance to the algorithm in \eqref{eq:kalman} which does not provide any privacy. Thus, we want to evaluate the sacrifice in output utility when using the privacy preserving algorithm. We thus simulate both algorithms solving the same problem and compare the results. We conduct the simulation on a laptop PC based on a Python implementation of the algorithms. We use $n=3$ computing parties and $t=1$. For the $\mathtt{sharing}$ algorithm we use mean value zero and variance 1000 for the Gaussian distributed shares. 

We use the RSE between the 
result from \proref{pro:RLS}, $\hat{x}_k^{\text{(priv)}}$, and \eqref{eq:kalman}, $\hat{x}_k$, which at time $k$ is defined as 
\begin{equation*}
\text{RSE}_k =  \sqrt{ (\hat{x}^{\text{(priv)}}_{k} - \hat{x}_k  )^2} , \quad \text{for } k = 1,2,\ldots. 
\end{equation*}

\begin{figure}
    \centering
    \input{plotKalmanRSE}
    \caption{RSE between simulated result from  \proref{pro:RLS} and \eqref{eq:kalman}.}  
    \label{fig:KalRSE}
\end{figure}
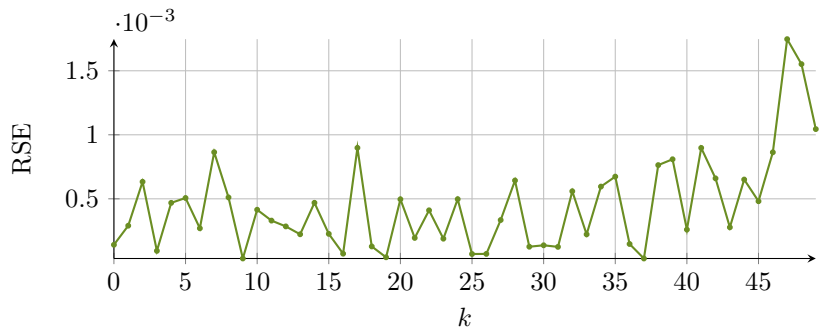

In \figref{fig:KalRSE} it is seen that the difference in result from the privacy preserving solution and the non-private solution lies around the third decimal. In comparison, the difference for the solution in \cite{secKalman} lies before the decimal point.

Regarding the complexities, as seen, \proref{pro:RLS} uses 12 multiplications and one inversion, which amounts to 27 interactive operations, independent of the dimension of the matrices. \cite{secKalman} does not provide the complexity for their solution, thus, we provide here an underestimation of the number of interactive operations which lies around $10M +l +1$, where $M$ is the dimension of the matrix $\v R$ in \eqref{eq:kalman} and $l$ is the number of bits used to represent the numbers (which in the simulations by \cite{secKalman} is at least 24 bits).

\section{Conclusion}\label{sec:conc}
The paper presents a real number secret sharing scheme that bypasses the requirements on integer shares and modular arithmetic which is used in state-of-the-art secure multiparty computation schemes. That the scheme does not use modular arithmetic, makes it very useful for computations directly on shares including division. The trade-off is that the proposed scheme is not perfectly secure, however, we show that the information leak can be upper bounded and demonstrate with examples how small the leak is. We see the proposed scheme with its high level accuracy and privacy properties and its low communication complexity as offering a relevant trade-off between between privacy of the distributed computations and practicality of the scheme.     
Numerical evaluations of the proposed scheme as well as simulations of the scheme to perform Kalman filtering with privacy preservation verify the theoretic results. 

\section*{Acknowledgement}
This work was supported by SECURE research project at Aalborg University.

\bibliography{mybibfile}

\end{document}

%% file: preamble.tex
\usepackage[linewidth=0.8pt]{mdframed}

\mdfdefinestyle{style}{%
  innermargin=-0pc,
  outermargin=-6pc,
  innerleftmargin=0pc,
  innerrightmargin=-6pc
}

\usepackage{array,booktabs}
\usepackage{framed}

\usepackage{amsmath, calc}
\usepackage{mathtools}
\usepackage{amssymb}
\usepackage{amsfonts}
\usepackage{subcaption}

\usepackage{algorithm}
\usepackage{algpseudocode}
\usepackage{amsthm}
\usepackage{hyperref}
\usepackage{xargs} 
\usepackage[dvipsnames,table,xcdraw]{xcolor}
\usepackage{pgfplots}
\usepackage{tikz}
\usetikzlibrary{positioning,arrows}
\usepackage{graphicx}
\usepackage{multirow}

\newcommand{\m}[1]{\ensuremath{{{#1}} }  }
\newcommand{\tub}[1]{ \ensuremath{ \{ {#1} \} }  }

\newcommand{\proref}[1]{Algorithm~\ref{#1}}

\newcommand{\figref}[1]{Fig. ~\ref{#1}}

\newcommand{\N}{{\ensuremath{\mathcal{N} }}}

\newcommand{\R}{{\ensuremath{\mathbb R} }}

\newcommand{\F}{{\ensuremath{\mathbb F_q} }}

\renewcommand{\v}[1]{\ensuremath{\boldsymbol{#1}}}

\newcommand{\cov}[1]{\ensuremath{\text{cov}\left({#1}\right)}}

\newcommand{\p}{{\ensuremath{\mathcal{P}} }}
\newcommand{\T}{{\ensuremath{\mathcal{T}} }}

\newcommand{\se}[2][p] {\ensuremath{#2[#1]}} 
\newcommand{\SE}[2][p] {\ensuremath{\{#2[#1] \}_{#1 \in \p} }}
\newcommand{\TE}[3][\T] {\ensuremath{\{#2[#3] \}_{#3 \in #1} }}

\DeclarePairedDelimiter\autobracket{(}{)}
\newcommand{\br}[1]{\autobracket*{#1}}

\newtheorem{prop}{Proposition}

\newtheorem{defi}{Definition}

\newtheorem{exa}{Example}

\makeatletter
\newcommand\fs@betterruled{%
  \def\@fs@cfont{\bfseries}\let\@fs@capt\floatc@ruled
  \def\@fs@pre{\vspace*{5pt}\hrule height.8pt depth0pt \kern2pt}%
  \def\@fs@post{\kern2pt\hrule\relax}%
  \def\@fs@mid{\kern2pt\hrule\kern2pt}%
  \let\@fs@iftopcapt\iftrue}
\makeatother

\makeatletter
\newsavebox{\mybox}
\newcounter{pro}

\newenvironment{pro}[1]{
\renewcommand{\ALG@name}{Algorithm}
\let\c@algocf\c@pro
\begin{algorithm}[#1]%
}%
{
\end{algorithm}}

\newcommand{\eqnum}{\refstepcounter{equation}\textup{\tagform@{\theequation}}}
\makeatother


%% file: plotG1.tex
\tikzset{%
    add1cm/.style={%
        execute at end picture={\path (current bounding box.north)--++(0,0.19cm);
        }
    },
    background rectangle/.style={fill=blue!30}
}

\definecolor{dgreen}{RGB}{107,142,35}
\definecolor{dpurple}{RGB}{0,0,139}
\definecolor{dorange}{RGB}{142,68,173}
\definecolor{red}{RGB}{178,34,34}
\definecolor{blue}{RGB}{241,196,15}

\begin{tikzpicture}[add1cm]

\begin{axis}
[
axis lines = left,
height = 4.5 cm,
width = \textwidth,
xlabel={$p$},
ylabel={\R},
xtick={0.5 , 0.65, 0.8 , 0.95, 1.1 , 1.25, 1.4 , 1.55, 1.7 , 1.85, 2},
grid=both,
every axis plot/.append style={thick},
legend style={at={(0.5,1.3)},
anchor=north,
font=\footnotesize},
legend columns=-1,
legend style={/tikz/every even column/.append style={column sep=0.5cm}},
every axis plot/.append style={thick}
]

\addplot [color = dgreen, only marks] coordinates {
(0.5,60.824576205346)
(0.65,91.07923335292908)
(0.8,134.68463733763565)
(0.95,197.32866303421173)
(1.1,286.7348290348702)
(1.25,413.32473380809324)
(1.4,590.8804918574372)
(1.55,837.207169880336)
(1.7,1174.7952229269029)
(1.8499999999999999,1631.4829305587368)
(2.0,2241.118833007725)
};
\addlegendentry{$s[p]$}

\end{axis}
\end{tikzpicture}

%% file: plotG4.tex
\tikzset{%
    add1cm/.style={%
        execute at end picture={\path (current bounding box.north)--++(0,0.19cm);
        }
    },
    background rectangle/.style={fill=blue!30}
}

\definecolor{dgreen}{RGB}{107,142,35}
\definecolor{dpurple}{RGB}{0,0,139}
\definecolor{dorange}{RGB}{142,68,173}
\definecolor{red}{RGB}{178,34,34}
\definecolor{blue}{RGB}{241,196,15}

\begin{tikzpicture}[add1cm]

\begin{axis}
[
axis lines = left,
height = 4.5 cm,
width = \textwidth,
xlabel={$p$},
ylabel={\R},
xtick={0.5 , 0.65, 0.8 , 0.95, 1.1 , 1.25, 1.4 , 1.55, 1.7 , 1.85, 2},
grid=both,
every axis plot/.append style={thick},
legend style={at={(0.5,1.3)},
anchor=north,
font=\footnotesize},
legend columns=-1,
legend style={/tikz/every even column/.append style={column sep=0.5cm}},
every axis plot/.append style={thick}
]

\addplot [color = dgreen, only marks] coordinates {
(0.5,33.504620401474895)
(0.65,32.468055313325124)
(0.8,24.00332981592596)
(0.95,28.023298312962414)
(1.1,135.99595643428378)
(1.25,632.7560888772248)
(1.4,2251.800332537602)
(1.55,6662.967981448535)
(1.7,17339.883679002676)
(1.8499999999999999,41017.268711151)
(2.0,90027.12074719659)
};

\end{axis}
\end{tikzpicture}

%% file: plotG2.tex
\tikzset{%
    add1cm/.style={%
        execute at end picture={\path (current bounding box.north)--++(0,0.19cm);
        }
    },
    background rectangle/.style={fill=blue!30}
}

\definecolor{dgreen}{RGB}{107,142,35}
\definecolor{dpurple}{RGB}{0,0,139}
\definecolor{dorange}{RGB}{142,68,173}
\definecolor{red}{RGB}{178,34,34}
\definecolor{blue}{RGB}{241,196,15}

\begin{tikzpicture}[add1cm]

\begin{axis}
[
axis lines = left,
height = 4.5 cm,
width = \textwidth,
xlabel={$p$},
ylabel={\R},
xtick={0, 0.5 , 0.65, 0.8 , 0.95, 1.1 , 1.25, 1.4 , 1.55, 1.7 , 1.85, 2},
grid=both,
every axis plot/.append style={thick},
legend style={at={(0.5,1.3)},
anchor=north,
font=\footnotesize},
legend columns=-1,
legend style={/tikz/every even column/.append style={column sep=0.5cm}},
every axis plot/.append style={thick}
]

\addplot [color = dgreen, only marks] coordinates {
(0.5,-466.5063877128687)
(0.65,393.6467938982267)
(0.8,747.0755365655176)
(0.95,602.6532621019152)
(1.1,163.2055872535697)
(1.25,-280.78744305822966)
(1.4,-457.4891224952931)
(1.55,-220.00385006059514)
(1.7,347.3251031434767)
(1.8499999999999999,822.6178271571639)
(2.0,340.1600064050799)
};
\addlegendentry{$s[p]$}

\addplot [color = red, only marks, mark = square, mark size = 3pt] coordinates {
(0.5,-466.5063877128687)
(0.65,393.6467938982267)
(0.95,602.6532621019152)
(1.4,-457.4891224952931)
(2.0,340.1600064050799)
};
\addlegendentry{$(x_j,y_j)_{j\in T}$ }


\end{axis}
\end{tikzpicture}

%% file: plotG3.tex
\tikzset{%
    add1cm/.style={%
        execute at end picture={\path (current bounding box.north)--++(0,0.19cm);
        }
    },
    background rectangle/.style={fill=blue!30}
}

\definecolor{dgreen}{RGB}{107,142,35}
\definecolor{dpurple}{RGB}{0,0,139}
\definecolor{dorange}{RGB}{142,68,173}
\definecolor{red}{RGB}{178,34,34}
\definecolor{blue}{RGB}{241,196,15}

\begin{tikzpicture}[add1cm]

\begin{axis}
[
axis lines = left,
height = 4.5 cm,
width = \textwidth,
xlabel={$p$},
ylabel={\R},
xtick={0, 0.5 , 0.65, 0.8 , 0.95, 1.1 , 1.25, 1.4 , 1.55, 1.7 , 1.85, 2},
grid=both,
every axis plot/.append style={thick},
legend style={at={(0.5,1.3)},
anchor=north,
font=\footnotesize},
legend columns=-1,
legend style={/tikz/every even column/.append style={column sep=0.5cm}},
every axis plot/.append style={thick}
]

\addplot [color = dgreen, only marks] coordinates {
(0.5,634.1955099739134)
(0.65,-336.87500989548863)
(0.8,2018.4757713258268)
(0.95,-1520.8835702512413)
(1.1,1016.8765149191023)
(1.25,794.4218683801591)
(1.4,-2471.2334214329717)
(1.55,92.32690359950065)
(1.7,1104.7551816262305)
(1.8499999999999999,-483.42095756605266)
(2.0,-202.3147855848074)
};

\addplot [color = red, only marks, mark = square, mark size = 3pt] coordinates {
(0.5,634.1955099739134)
(0.65,-336.87500989548863)
(0.8,2018.4757713258268)
(0.95,-1520.8835702512413)
(1.25,794.4218683801591)
(1.4,-2471.2334214329717)
(1.55,92.32690359950065)
(1.7,1104.7551816262305)
(1.8499999999999999,-483.42095756605266)
(2.0,-202.3147855848074)
};


\end{axis}
\end{tikzpicture}

%% file: plotExaMult.tex
\tikzset{%
    add1cm/.style={%
        execute at end picture={\path (current bounding box.north)--++(0,0.19cm);
        }
    },
    background rectangle/.style={fill=blue!30}
}

\definecolor{dgreen}{RGB}{107,142,35}
\definecolor{dpurple}{RGB}{0,0,139}
\definecolor{dorange}{RGB}{142,68,173}
\definecolor{red}{RGB}{178,34,34}
\definecolor{blue}{RGB}{241,196,15}

\begin{tikzpicture}[add1cm]

\begin{axis}
[
axis lines = left,
height = 4.5 cm,
width = 0.75\textwidth,
ylabel={\R},
xticklabels={ $1$, $2$,$3$,$d$,$e$},
xtick={1,...,5},
grid=both,
every axis plot/.append style={thick},
legend style={at={(0.5,1.3)},
anchor=north,
font=\footnotesize},
legend columns=-1,
legend style={/tikz/every even column/.append style={column sep=0.5cm}},
every axis plot/.append style={thick}
]


\addplot [color = white] coordinates {
(0,0)
};
\addlegendentry{ }

\addplot [color = dpurple, only marks, mark = diamond] coordinates {
(1.0,618.71381159)
(2.0,-659.22167926)
(3.0,-1449.87972773)
};
\addlegendentry{$s_1[p]$}

\addplot [color = dgreen, only marks, mark = diamond] coordinates {
(1.0,-1032.44269697)
(2.0,-289.51599489)
(3.0,1404.85865413)
};
\addlegendentry{$s_2[p]$}

\addplot [color = red, only marks] coordinates {
(4,1196.40101103)
};
\addlegendentry{$d$}

\addplot [color = blue, only marks] coordinates {
(5,-340.48298834)
};
\addlegendentry{$e$}

\end{axis}
\end{tikzpicture}

%% file: plotAccuracy.tex
\tikzset{%
    add1cm/.style={%
        execute at end picture={\path (current bounding box.north)--++(0,0.19cm);
        }
    },
    background rectangle/.style={fill=blue!30}
}

\definecolor{dgreen}{RGB}{107,142,35}
\definecolor{dpurple}{RGB}{0,0,139}
\definecolor{dorange}{RGB}{142,68,173}
\definecolor{red}{RGB}{178,34,34}
\definecolor{blue}{RGB}{241,196,15}
\definecolor{orang}{RGB}{179,89,0}

\begin{tikzpicture}[add1cm]

\begin{semilogyaxis}
[
axis lines = left,
height = 7.5 cm,
width = \textwidth,
xlabel={$\sigma^2_{Y}$},
ylabel={RSE},
x grid style={white!69.01960784313725!black},
grid=both,
every axis plot/.append style={thick},
legend columns=-1,
every axis plot/.append style={thick},
legend style={column sep=5pt,
            draw=none,
              font=\footnotesize,
              align = left,
              at={(0.5,1.1)},anchor=north}
]



\addplot [color = dpurple,dashed, line width = 2pt] coordinates {
(1,2.1946098271996562e-11)
(21,4.0353332231291006e-10)
(41,7.818376701607121e-10)
(61,1.28296024470842e-09)
(81,1.9532489758944392e-09)
(101,2.6171240996575305e-09)
(121,1.863663801060511e-09)
(141,2.8630219883751806e-09)
(161,3.806352371071853e-09)
(181,5.779068574796042e-09)
(201,4.498556478438332e-09)
(221,3.0500898873242476e-09)
(241,4.796711330357084e-09)
(261,5.216181477862847e-09)
(281,4.999082943157873e-09)
(301,5.993268654691519e-09)
(321,5.800830731317319e-09)
(341,5.3220987261681784e-09)
(361,9.56310373112501e-09)
(381,6.478360958794838e-09)
(401,6.874573053039512e-09)
(421,7.047241439295249e-09)
(441,1.1181740049437394e-08)
(461,7.252649627531582e-09)
(481,8.051369206896197e-09)
(501,1.3284355473430764e-08)
(521,1.3162612866324252e-08)
(541,9.110580201365792e-09)
(561,1.5830344395340036e-08)
(581,1.2663279669311578e-08)
(601,1.1343023009757758e-08)
(621,1.1542536118014368e-08)
(641,1.2406177631874016e-08)
(661,1.303272085806384e-08)
(681,1.3493599393399337e-08)
(701,1.1707058469312414e-08)
(721,1.9588539217352264e-08)
(741,1.519186483278645e-08)
(761,2.490173866220857e-08)
(781,1.4255162197329695e-08)
(801,1.3489073520389638e-08)
(821,2.6885015707023285e-08)
(841,1.7674296373115795e-08)
(861,1.998307812556277e-08)
(881,1.1800258050698175e-08)
(901,1.525975008576097e-08)
(921,2.1103924510157413e-08)
(941,1.8417426019823323e-08)
(961,2.019365913596971e-08)
(981,2.0317181252949013e-08)
};
\addlegendentry{$\mathtt{recon}$}

\addplot [color = orang, line width = 1pt] coordinates {
(1,6.288296106049529e-11)
(21,6.37498587252594e-10)
(41,1.368736093354528e-09)
(61,1.5397351660340063e-09)
(81,2.3608892263382586e-09)
(101,2.998716723823236e-09)
(121,4.6932614594652475e-09)
(141,5.119517183516109e-09)
(161,5.167444712128599e-09)
(181,4.756833718033704e-09)
(201,5.685626547347056e-09)
(221,6.0651357358665335e-09)
(241,6.730672197363674e-09)
(261,9.564215872615023e-09)
(281,1.034524217402577e-08)
(301,1.1411471163569331e-08)
(321,8.241556699317698e-09)
(341,1.100477476256856e-08)
(361,1.3011158088716001e-08)
(381,1.1250564142528673e-08)
(401,1.1084750326517678e-08)
(421,1.370314308246634e-08)
(441,1.6273681353595748e-08)
(461,1.3073443767552816e-08)
(481,1.5993599049579643e-08)
(501,2.000973047699972e-08)
(521,1.6329586216556892e-08)
(541,1.914877344688648e-08)
(561,2.0367829236533907e-08)
(581,2.4500762165757806e-08)
(601,1.7701902947919733e-08)
(621,2.7795539239150457e-08)
(641,2.1664650660113694e-08)
(661,2.096468293188991e-08)
(681,2.3811640375015486e-08)
(701,2.429008297610835e-08)
(721,3.0324140425364024e-08)
(741,3.335609441990073e-08)
(761,2.8752393177455816e-08)
(781,2.11448826803462e-08)
(801,2.2395273333586374e-08)
(821,3.25454115568391e-08)
(841,2.276580559623653e-08)
(861,2.485196965551495e-08)
(881,2.4294693830029245e-08)
(901,3.2357495456381e-08)
(921,3.444007717234854e-08)
(941,2.5858710017701014e-08)
(961,4.320553514958192e-08)
(981,2.805811810446812e-08)
};
\addlegendentry{$\mathtt{add}$}

\addplot [color = dgreen, dash pattern=on 1pt off 3pt on 3pt off 3pt, line width = 2pt] coordinates {
(1,2.0376778184072464e-06)
(21,2.131509011178423e-06)
(41,2.1471636864589527e-06)
(61,2.8137167441855125e-06)
(81,4.0120303344792775e-06)
(101,3.007850407357182e-06)
(121,3.892016002282616e-06)
(141,5.95324415144205e-06)
(161,5.642671869736659e-06)
(181,5.759464954735449e-06)
(201,5.437030824850808e-06)
(221,5.34597906835188e-06)
(241,9.502209893526015e-06)
(261,8.831263805859635e-06)
(281,9.428030634808238e-06)
(301,8.604503293554444e-06)
(321,1.1058493257678492e-05)
(341,9.128775971021242e-06)
(361,1.1236374573400099e-05)
(381,1.0093628764025198e-05)
(401,1.644914573944334e-05)
(421,1.2559959679947497e-05)
(441,1.150069338564208e-05)
(461,1.027776510682088e-05)
(481,1.3665170611147914e-05)
(501,1.2558693113078334e-05)
(521,1.6699897673788656e-05)
(541,1.6452621632652154e-05)
(561,1.5972557114309894e-05)
(581,1.7267311836803855e-05)
(601,1.8504571951325488e-05)
(621,1.619150271835679e-05)
(641,1.7597142459635506e-05)
(661,9.966182303742244e-06)
(681,2.3331312956997862e-05)
(701,1.9627325559667953e-05)
(721,2.3061135211719376e-05)
(741,2.1598597988941038e-05)
(761,2.5784873475913627e-05)
(781,2.6969393761646643e-05)
(801,1.5714151129486708e-05)
(821,4.050682397746641e-05)
(841,2.6146849559154362e-05)
(861,2.7645913751257466e-05)
(881,2.6181312018707105e-05)
(901,3.03654534081943e-05)
(921,2.0415901256001236e-05)
(941,2.577056278823875e-05)
(961,2.7061073167260476e-05)
(981,3.154416114739433e-05)
};
\addlegendentry{$\mathtt{mult}$}

\addplot [color = blue, line width = 1pt] coordinates {
(1,3.319430375015031e-10)
(21,1.89867669364574e-10)
(41,2.0841507325508602e-10)
(61,2.1273057621007396e-10)
(81,2.591614389757524e-10)
(101,3.497072328939588e-10)
(121,5.965933116969246e-10)
(141,3.4357419603781027e-10)
(161,2.2100400148827858e-10)
(181,2.502831736306277e-10)
(201,4.329194663865721e-10)
(221,3.519033087151513e-10)
(241,2.795598542382205e-09)
(261,3.8846888122945346e-10)
(281,6.373574237827917e-10)
(301,4.917086246503111e-10)
(321,3.177722041058928e-10)
(341,5.169624808787354e-10)
(361,7.075981586779356e-10)
(381,7.008711558342817e-10)
(401,3.6775322898030537e-10)
(421,4.962320285240019e-10)
(441,4.4368433177455203e-10)
(461,4.3396847060339637e-10)
(481,6.780002081985259e-10)
(501,8.323364486528817e-10)
(521,3.0575234372109963e-10)
(541,6.038439151478947e-10)
(561,6.312432790367417e-10)
(581,9.228096628510763e-10)
(601,4.219646365655194e-10)
(621,4.954607485396778e-10)
(641,5.872306946264061e-10)
(661,1.066601772459208e-09)
(681,2.6693249541298592e-09)
(701,6.275588607107529e-10)
(721,7.35784835559361e-10)
(741,5.448732609547591e-10)
(761,7.377100638694678e-10)
(781,5.361750543553256e-10)
(801,8.208057644676359e-10)
(821,9.409111903502243e-10)
(841,7.982000238548892e-10)
(861,9.499831299386585e-10)
(881,6.440886657355094e-10)
(901,1.2403781157743765e-09)
(921,7.556521744445099e-10)
(941,6.068111430002077e-10)
(961,7.354246434054801e-10)
(981,1.5767780384146057e-09)
};
\addlegendentry{$\mathtt{inv}$}

\end{semilogyaxis}
\end{tikzpicture}

%% file: plotMI1share.tex
\tikzset{%
    add1cm/.style={%
        execute at end picture={\path (current bounding box.north)--++(0,0.19cm);
        }
    },
    background rectangle/.style={fill=blue!30}
}

\definecolor{dgreen}{RGB}{107,142,35}
\definecolor{dpurple}{RGB}{0,0,139}
\definecolor{dorange}{RGB}{142,68,173}
\definecolor{red}{RGB}{178,34,34}
\definecolor{blue}{RGB}{241,196,15}
\definecolor{orang}{RGB}{179,89,0}

\begin{tikzpicture}[add1cm]

\begin{semilogyaxis}
[
axis lines = left,
height = 7 cm,
width = \textwidth,
xlabel={$\sigma^2_{Y}$},
ylabel={MI [bits]},
x grid style={white!69.01960784313725!black},
grid=both,
legend columns=1,
every axis plot/.append style={thick},
legend style={column sep=5pt,
            draw=none,
              font=\footnotesize}
]



\addplot [color = dpurple, dash pattern=on 1pt off 3pt on 3pt off 3pt, line width = 2pt] coordinates {
(1,0.06088872797484961)
(21,0.00027856120329730063)
(41,8.670968966731252e-05)
(61,5.201068558623234e-05)
(81,4.494569639032164e-05)
(101,3.228935991367976e-05)
(121,2.9473475303802844e-05)
(141,2.7415966749728682e-05)
(161,2.707632942613003e-05)
(181,2.4687797836016047e-05)
(201,2.0804761572286167e-05)
(221,1.6207968704797793e-05)
(241,1.7323765764878374e-05)
(261,1.5625034557800178e-05)
(281,1.35182350798857e-05)
(301,1.3352015884926516e-05)
(321,1.4832951152392582e-05)
(341,9.69721795755163e-06)
(361,1.1082236550947755e-05)
(381,1.0945926889895929e-05)
(401,9.355053995356854e-06)
(421,9.3588182236104e-06)
(441,1.0276303813725462e-05)
(461,9.560372553565344e-06)
(481,1.0734581132965105e-05)
(501,8.116436942344763e-06)
(521,7.3361169657104594e-06)
(541,8.193839998078545e-06)
(561,8.7689924466261e-06)
(581,5.900344065672414e-06)
(601,6.941014031625059e-06)
(621,8.018504216416034e-06)
(641,7.781951443703861e-06)
(661,5.8986868179999875e-06)
(681,5.8760838749805085e-06)
(701,5.507176879007147e-06)
(721,5.516364133058005e-06)
(741,8.173392608554763e-06)
(761,6.48302232132636e-06)
(781,4.276758136860792e-06)
(801,4.590358870792954e-06)
(821,5.214813279614816e-06)
(841,3.360073340061831e-06)
(861,4.266876001572939e-06)
(881,3.887682883316756e-06)
(901,4.743704477903066e-06)
(921,3.1574783167798158e-06)
(941,5.359556956001654e-06)
(961,4.035665323200987e-06)
(981,4.2811293921829475e-06)
};
\addlegendentry{$I(S ; S[1])$}

\addplot [color = dgreen, dashed, line width = 2pt] coordinates {
(1,0.17868011370904896)
(21,0.006471856864617393)
(41,0.003618557503851605)
(61,0.002203456459647839)
(81,0.003786200763137231)
(101,0.0013886992622258276)
(121,0.0019570284369655154)
(141,0.0014601647278732345)
(161,0.001036295839774084)
(181,0.001005672392118697)
(201,0.0007185878691549874)
(221,0.0007899067575709751)
(241,0.000822212118961545)
(261,0.0009736367419748149)
(281,0.0005818243223993136)
(301,0.0006355822652145093)
(321,0.0007057444202371244)
(341,0.0006258850514075221)
(361,0.0005406762889523975)
(381,0.000910922401078551)
(401,0.00039289620959323204)
(421,0.00038627354910232727)
(441,0.0003861504063252141)
(461,0.0004663485204659423)
(481,0.00034778015488370786)
(501,0.0002497212917396041)
(521,0.00029223336515798336)
(541,0.0003844022353095511)
(561,0.00028896162679018287)
(581,0.0002774725699720193)
(601,0.0003323779192874099)
(621,0.00031911243041349735)
(641,0.00023455144638347746)
(661,0.0002841903394832457)
(681,0.0003555944214002693)
(701,0.00030131687287223487)
(721,0.00022838280311020752)
(741,0.0001777276577470843)
(761,0.00023295838615752017)
(781,0.0002984153484916874)
(801,0.00022149767012564326)
(821,0.0002111224562295888)
(841,0.0001596805372014387)
(861,0.0002044102126271241)
(881,0.00023690855714846747)
(901,0.00018572468165189092)
(921,0.00011550608548731134)
(941,0.0001869521067337132)
(961,0.00018678437716005192)
(981,0.00018293925232228502)
};
\addlegendentry{$I(S ; S[1],\ldots, S[t])$}

\addplot [color = orang, line width = 1pt] coordinates {
(1,0.13540395827506557)
(21,0.006695081797507782)
(41,0.0027040459938274353)
(61,0.0025948506307756245)
(81,0.0020281714565049923)
(101,0.00212096982836016)
(121,0.0010911536018221569)
(141,0.001009542969977133)
(161,0.0007087102189711913)
(181,0.001938059609297511)
(201,0.0010342977988662483)
(221,0.001080297150550109)
(241,0.0011961215929591163)
(261,0.000568410558182677)
(281,0.0005234705298946096)
(301,0.0005010464678542803)
(321,0.0005201043761937285)
(341,0.0008918489081003144)
(361,0.00027397273782469255)
(381,0.0003311293485401734)
(401,0.0004367568404029498)
(421,0.0003643254969671261)
(441,0.00027162500378409506)
(461,0.00033271117526282977)
(481,0.0003595411164667439)
(501,0.0002896046382116424)
(521,0.00048485832916639993)
(541,0.00034095576274694394)
(561,0.00029262980270488014)
(581,0.0002174284943288285)
(601,0.00033860108809392387)
(621,0.00020070599990930306)
(641,0.0003031394759051409)
(661,0.0002747726896666336)
(681,0.0003390287426114469)
(701,0.0003070673950537639)
(721,0.0004733136903772106)
(741,0.00019208886089643328)
(761,0.00022743163219388407)
(781,0.0003118002415760657)
(801,0.00018627696860988862)
(821,0.0001793609863130996)
(841,0.0003079850708218146)
(861,0.00028176817550701116)
(881,0.0006525923585380156)
(901,0.000415765434132993)
(921,0.00014375188062956567)
(941,0.00025965644861770444)
(961,0.00015150224106037059)
(981,0.00022876194446581622)
};
\addlegendentry{$I(S ; S[1],\ldots, S[t], S+R_1)$}

\end{semilogyaxis}
\end{tikzpicture}

%% file: plotKalmanRSE.tex
\tikzset{%
    add1cm/.style={%
        execute at end picture={\path (current bounding box.north)--++(0,0.19cm);
        }
    },
    background rectangle/.style={fill=blue!30}
}

\definecolor{dgreen}{RGB}{107,142,35}
\definecolor{dpurple}{RGB}{0,0,139}
\definecolor{dorange}{RGB}{142,68,173}
\definecolor{red}{RGB}{178,34,34}
\definecolor{blue}{RGB}{241,196,15}
\definecolor{orang}{RGB}{179,89,0}

\begin{tikzpicture}[add1cm]

\begin{axis}
[
axis lines = left,
height = 4.5 cm,
width = 0.9\textwidth,
xlabel={$k$},
ylabel={RSE},
grid=both,
every axis plot/.append style={thick},
legend columns=1,
every axis plot/.append style={thick},
legend style={column sep=5pt,
            draw=none,
              font=\footnotesize,
              align = left,
              at={(0.8,1)},anchor=north}
]



\addplot [color = dgreen, mark = *, mark size = 0.7pt] coordinates {
(0,0.00014100846370590503)
(1,0.0002901094916945568)
(2,0.0006330677984597299)
(3,9.359412400833378e-05)
(4,0.00046857930943300374)
(5,0.000506523205030307)
(6,0.00026977563691144013)
(7,0.0008641778717481952)
(8,0.0005110090267550493)
(9,3.405509700105114e-05)
(10,0.0004147634150641899)
(11,0.00032990321751746876)
(12,0.00028455355151746353)
(13,0.00022302402348506334)
(14,0.00046899199897332267)
(15,0.0002262624969571192)
(16,7.1554276319441e-05)
(17,0.0008980625689015698)
(18,0.00012801504448084478)
(19,4.3338365087353115e-05)
(20,0.0004965915593241332)
(21,0.00019471867155274047)
(22,0.000410224626492095)
(23,0.0001887139049672193)
(24,0.000497459985819404)
(25,6.896407034950869e-05)
(26,6.982761890550471e-05)
(27,0.00033427739867386563)
(28,0.0006430547396902853)
(29,0.00012597035641537246)
(30,0.0001369797889223534)
(31,0.00012465773935255164)
(32,0.0005591141453193949)
(33,0.0002223467873172444)
(34,0.0005955839729842594)
(35,0.0006734672844797007)
(36,0.0001471615384038838)
(37,3.388255215841518e-05)
(38,0.0007633712725683139)
(39,0.0008088017963383543)
(40,0.0002594721245228193)
(41,0.0008976256248480752)
(42,0.0006596005941368333)
(43,0.0002773858598292822)
(44,0.0006505998156696169)
(45,0.00048106319872709413)
(46,0.0008632294286312625)
(47,0.0017469962362607339)
(48,0.0015518539061558645)
(49,0.0010445591196192039)
};

\end{axis}
\end{tikzpicture}